\documentclass[pre,twocolumn,nofootinbib,twoside,showpacs,superscriptaddress,tightenlines,floatfix,aps]{revtex4-2}
\usepackage{dcolumn}
\usepackage{lipsum}
\usepackage{graphicx,amssymb,amsmath,epsf,bm}
\usepackage[utf8x]{inputenc}
\usepackage[T1]{fontenc}
\usepackage[normalem]{ulem}
\usepackage{color}
\usepackage{mathtools}
\usepackage{dsfont}
\usepackage{nicefrac}
\usepackage{hyperref}
\hypersetup{
    allcolors=blue,
    colorlinks=true,
    bookmarks=true,
    pdfpagemode=FullScreen,
}

\usepackage{ulem}
\usepackage{tikz}
\usepackage{bm}
\usepackage{txfonts}
\usepackage{mathrsfs}
\usepackage{amsmath}
\usepackage{xcolor}
\usepackage{upgreek}

\definecolor{nred}{RGB}{224,0,0}
\definecolor{nblue}  {RGB}{28,130,185}
\definecolor{dgreen} {RGB}{38,238,21}
\definecolor{norange}{RGB}{230,120,20}

\newcommand{\Tr}{{\rm Tr}}

\begin{document}

\title{Generalization of the Central Limit Theorem to Critical Systems: Revisiting Perturbation Theory}

\author{Sankarshan Sahu }
\affiliation{Sorbonne Universit\'e, CNRS, Laboratoire de Physique
Th\'eorique de la Mati\`ere Condens\'ee, LPTMC, 75005 Paris, France}
\author{Bertrand Delamotte}
\affiliation{Sorbonne Universit\'e, CNRS, Laboratoire de Physique
Th\'eorique de la Mati\`ere Condens\'ee, LPTMC, 75005 Paris, France}
\author{Adam Ran\c con}
\affiliation{Univ. Lille, CNRS, UMR 8523 – PhLAM – Laboratoire de Physique des Lasers Atomes et Mol\'ecules, F-59000 Lille, France}
\date{\today}
\begin{abstract}
The Central Limit Theorem does not hold for strongly correlated stochastic variables, as is the case for statistical systems close to criticality. Recently, the calculation of the probability distribution function (PDF) of the magnetization mode has been performed with the functional renormalization group in the case of the three-dimensional Ising model [Balog et al., Phys. Rev. Lett. {\bf 129}, 210602 (2022)]. It has been shown in that article that there exists an entire family of universal PDFs parameterized by $\zeta=\lim_{L,\xi_\infty\rightarrow\infty} L/\xi_\infty$ which is the ratio of the system size $L$ to the bulk correlation length $\xi_{\infty}$ with both the thermodynamic limit and the critical limit being taken simultaneously. We show how these PDFs or, equivalently, the rate functions which are their logarithm, can be systematically computed perturbatively in the $\epsilon=4-d$ expansion. We determine the whole family of universal PDFs and show that they are in good qualitative agreement with Monte Carlo data.   Finally, we conjecture on how to significantly improve the quantitative agreement between the one-loop and the numerical results.
\end{abstract}


\maketitle

\section{Introduction}
The physics of many-body systems has seen many important advances over the twentieth century. For both particle and statistical physics, field theory has provided a unified formalism, particularly when the physics involves length scales much larger than the microscopic scales at which the models are formulated. This is the usual case in particle physics where the ultraviolet (UV) cutoff $\Lambda$ of the theory is in general unknown, that is, it is so large that its effects on low-energy physics are almost invisible \cite{ZinnForChildren}. This is also the case in statistical mechanics when the system is close to a second-order phase transition. In this case, the correlation length is very large compared to the microscopic scale -- a lattice spacing, an inter-molecular distance, etc -- and here again, the long-distance physics is almost insensitive to its UV cutoff. In both cases, renormalization plays a crucial role. Historically, renormalization was invented to eliminate UV divergences appearing at all orders of perturbation theory.  When the theory is either renormalizable or superrenormalizable, it consists of a reparametrization, called renormalization, of coupling constants, masses, and fields leading to finite correlation functions. In the renormalization process, all reference to this UV scale is eliminated, and this is precisely what makes it possible to get rid of the UV cutoff. It is replaced by the scale where the renormalized couplings are defined. As a consequence, these couplings become scale-dependent, and their evolution with length or energy scale is given by the renormalization group (RG) equations. Renormalization is thus one of the cornerstones of the study of many-body systems.

A second cornerstone of the study of many-body systems is the Central Limit Theorem (CLT) which states that for a set of $n$ independent and identically distributed (iid) random variables, the probability distribution function (PDF) of their properly normalized sum is a Gaussian when $n\to\infty$, provided that their mean and variance exist \cite{FellerBook, Botet_book}. This theorem can easily be generalized to the case of weakly correlated random variables by considering groups of correlated variables that, on scales larger than the correlation length, are effectively uncorrelated and identically distributed \cite{Dedecker2007,Botet_book}. In this case, the variance of the individual degrees of freedom is replaced by the susceptibility of the system, and the CLT still holds.

The two cornerstones described above are in fact two facets of the same key concept of many-body systems: Universality.
 For a many-body system, universality is most often described as the complete loss of memory of microscopic details at the macroscopic level, the idea being that the microscopic details of a physical system are averaged out at large scales and that what only matters for the long-distance physics is the tensorial nature of the order parameter -- for the symmetry group of the system -- and the space dimensionality. It is often stated that universality requires scale invariance which, existing only at criticality, makes universality the main and most striking feature of second-order phase transitions. This is however too restrictive. Universality in the above sense is in fact at work everywhere in physics, and is precisely what makes physics possible; otherwise, for a gas of weakly interacting molecules, instead of using the ideal gas law, we would have to solve the dynamics of all the quarks and electrons in all the atoms of the gas, not to mention their quantum gravitational interactions. Although below the molecular scale, the physics of the elementary degrees of freedom of the gas -- electrons, quarks, etc. -- is non-trivial, beyond this scale, it is the CLT that enables us to overcome this difficulty and is, therefore, the most common manifestation of universality. 
 
 The difficulty with criticality is that all degrees of freedom are long-range correlated and, consequently, the CLT does not apply. Therefore, the real surprise with universality is not that it is specific to criticality, but that, on the contrary, criticality shows universality. One of the aims of the present work is to show by perturbative means that the CLT can be generalized to critical systems and that the PDF of the sum of the system's stochastic variables is no longer Gaussian.\footnote{In the following, we call for short PDF the PDF of the properly normalized sum of the system's stochastic variables.} More precisely, using the $\epsilon$-expansion, we show that at criticality there are infinitely many such PDF depending on the way the critical limit, $T\to T_c$, and the thermodynamic limit, $L\to\infty$, are taken. Here $T$ is the system's temperature, $T_c$ its critical temperature and $L$ the (linear) system size. We also show that all these PDFs, indexed by the ratio $\zeta=\lim_{L,\xi_\infty\rightarrow\infty}L/\xi_{\infty}$ where $\xi_{\infty}$ is the correlation length of the infinite size system, are universal. Although universal, PDFs depend on the boundary conditions \cite{Binder1981a}. While some quantities, such as critical exponents, remain independent of boundary conditions, finite-size quantities generally do not. At criticality, long-range correlations cause effects of the boundaries to persist even as $L\to\infty$. Consequently, for a given $\zeta$, there exists an infinite family of PDFs, each corresponding to different boundary conditions. We focus here on periodic boundary conditions only.

Clearly, renormalization has to do with universality because the very fact that the UV cutoff can be eliminated in a renormalizable theory while maintaining fixed the physics at large scales means that microscopic details of the model do not play any crucial role at scales much larger than the cutoff. 
The speculation of the relationship between renormalization group and probability theory goes back at least as far as  1973 when Bleher and Sinai \cite{Bleher1973,Bleher1987} hinted towards it in the context of hierarchical models, which are nontrivial statistical systems specially designed such that they can be solved exactly with the RG. This idea was extended by Jona-Lasinio \cite{JonaLasino1975}, where he further established the connection between limiting theorems in probability theory and RG, e.g. see \cite{Cassandro1978,Sinai1998,Koralov2012}. 

 In the language of the RG, the relationship goes as follows: The coarse-graining procedure at the heart of Wilson's RG consists in integrating out fluctuations of wavenumber $\vert q\vert \in\ ]k,\Lambda] $ and obtaining the effective Hamiltonian ${\cal H}_k$ for the modes of wavenumber smaller than $k$. Therefore, when all modes have been integrated out except the zero momentum mode, i.e. the sum of all variables as in the CLT, the PDF of this sum is given by $\exp(-{\cal H}_{k=0})$ up to a normalization factor. 
 The flow of probability measures is thus directly related to the flow of the coupling constants in the potential part of  ${\cal H}_k$. It has already been shown in \cite{Balog2022} that when the critical limit is taken first and only then the thermodynamic limit, the log of the PDF is close to (albeit slightly different from) the fixed-point potential, reinstating the connection between RG and probability theory for correlated random variables.

The PDF calculation is not only interesting from a conceptual point of view but also from a pragmatic one. Obviously,  universal quantities are much easier to compare between theory and experiments than nonuniversal quantities.\footnote{It should be noted that nonuniversal quantities have also been successfully calculated for some simple models such as the 3D Ising model \cite{Machado2010}, quantum lattice models \cite{Rancon2011,Rancon2014b} and reaction-diffusion systems \cite{Canet2004} using a modern nonperturbative version of Wilson's RG. However, this remains a challenge in general and is mostly impossible by perturbative or conformal bootstrap means.} Critical exponents are the most famous and most studied universal quantities. Six critical exponents rule the leading scaling behavior of the thermodynamic quantities,  among which only two are independent. They have been computed with striking accuracy (in both the perturbative as well as nonperturbative setting). However, they remain difficult to compute numerically and measure experimentally, not least because, in most cases, scaling only takes place over a finite interval of the control parameter, there is disorder and/or impurities, etc. Moreover, in many cases, only one or two of these exponents can be measured. On the theoretical side, an accurate perturbative determination of these exponents is difficult because it requires going to rather large orders of perturbation theory and performing resummations. Conformal bootstrap is free of these difficulties but has been implemented so far on only a limited number of models. The same holds true for the functional RG. Correction to scaling exponents and amplitude ratios are also universal \cite{Pelissetto2002}. They have been computed perturbatively for rather simple models, e.g. in O($N$) models, with similar difficulties. Even though in principle there are infinitely many of these universal quantities, what we have in the end are, most of the time, only two numbers to characterize a universality class.

Of course, universal functions also exist and in principle provide a better characterization of universality classes. However, in practice, they are often both difficult to compute accurately and to measure experimentally with the prime example being that of the structure factor which is the universal part of the correlation function \cite{Benitez2012}. The PDF is another obvious example of a universal function. A lot of efforts have been made in the past decades to devise such a PDF for strongly correlated systems most of which are rather unsystematic and riddled with mathematical discrepancies \cite{Bruce1979,Rudnick1985,Eisenriegler1987,Hilfer1993,Hilfer1995,Esser1995,Chen1996,Bruce1997,Rudnick1998}. Recently, the PDF has been computed for the three-dimensional Ising model using a modern version of Wilson's RG called either the functional RG or the nonperturbative RG (NPRG) \cite{Balog2022}. 

We show in this article, harnessing the power of the RG, that the PDF can be calculated perturbatively although in essence this calculation remains functional. A first perturbative calculation of the PDF in the Ising model was done long ago by Eisenriegler and Tomaschitz \cite{Eisenriegler1987}. In our opinion, this article has not received the attention it deserves, probably because it is not an easy read and the comparison with numerical simulations was not easy to make at the time (we disagree with their comparison shown in their Fig.~1, see below), and the tail of the PDF was unreachable by Monte Carlo simulations.

In this article, we present a systematic method, hopefully in a pedagogical way, for deriving at one loop the entire family of PDFs, indexed by the ratio $\zeta$ in the Ising and O($N$) models (with $N=1$ corresponding to the Ising model).
 As in \cite{Eisenriegler1987}, we show that these PDFs do not exhibit the correct large field behavior which calls for an improvement of the tail of the PDFs. 
Our work  generalizes \cite{Eisenriegler1987} by computing the entire $\zeta$-dependent family of universal PDFs and rate functions. It shows that the PDFs can be unimodal or bimodal depending on $\zeta$. Using RG arguments, it is shown how to perform improvements of the tails of the rate functions for the Ising model for all values of $\zeta$ and how, using the Principle of Minimal Sensitivity, these improvements can be optimized. By comparing with Monte Carlo data, we also show that our results are qualitatively correct for all values of $\zeta$, including the negative values corresponding to the approach to criticality from the low-temperature phase. However, the agreement is not quantitative. Finally, we show that if we rescale the entire family of rate functions by one number only, the agreement between the Monte Carlo and the improved rate functions becomes excellent for all values of the field and for all values of $\zeta$. We therefore formulate the conjecture that the main inaccuracy of the one-loop approximation is concentrated in that one number which is the overall scale of the rate functions. 
The manuscript is organized as follows. The general formalism is described in Sec.~\ref{sec:field_theory}, and two approaches to compute the rate function at one loop are given in Secs.~\ref{sec:1loop} and \ref{sec:alternative}.
The RG improvement to correctly capture the large field behavior is described in Sec.~\ref{improvement}, while Sec.~\ref{sec:lowT} discusses a generalization of our calculation to approach to criticality from the low-temperature phase. Finally, we compare our results to Monte Carlo simulations in Sec.~\ref{MC} and discuss our conjecture in Sec.~\ref{conjecture}. Our conclusions are given in Sec.~\ref{conclusion}.

\section{The field theoretic formalism \label{sec:field_theory}}

Let us first recall what the PDF is in the case of weakly correlated random variables. We consider in the following the ferromagnetic Ising model or more generally the O($N$) model. 

We therefore consider random variables $\hat\sigma_{i}$ that are $N$-component unit vectors defined on each site $i$ of a hypercubic lattice in $d$ dimensions with periodic boundary conditions of linear size $L$ and lattice spacing $a$. Their joint probability distribution is given by the Boltzmann law: $\exp(-{\cal H}/k_BT)/Z$ with Hamiltonian 
\begin{equation}
    {\cal H}= -J\sum_{\langle ij\rangle} \hat\sigma_{i}\cdot\hat\sigma_{j},
\end{equation}
where $J>0$. The system is said to be weakly correlated if the correlation matrix $G_{ij}\propto \langle \hat\sigma_{i}\hat\sigma_{j}\rangle$ decays sufficiently fast at large distance such that  $\sum_i G_{ij}$ is finite. Typically, when $L\to\infty$ and $T>T_c$ where $T_c$ is the critical temperature of the infinite volume system, $G_{ij}$ decays exponentially with the distance between sites $i$ and $j$ with a characteristic length scale given by the (bulk) correlation length $\xi_\infty$. Physically, the picture becomes that of finite clusters of spins of size  $\xi_\infty$ which are strongly correlated within the cluster, with each cluster being independent of each other \cite{Bouchaud1990}. 

What we are interested in is the PDF  of the total normalized sum of spins defined as $\hat{s}=L^{-d}\sum_{i}\hat\sigma_{i}$, the average of which is the magnetization per spin $m=\langle\hat{s}\rangle$. We represent this quantity by $P(\hat{s}=s)$. The fluctuations of $\hat{s}$ are measured by the quantity $\langle\hat{s}^{2}\rangle=L^{-d}\chi$ with $\chi$ the magnetic susceptibility. In the limit of infinite volume ($L\rightarrow\infty$)  and for a weakly correlated system, i.e. for $T>T_{c}$, one recovers the Central Limit Theorem, i.e. $P(\hat{s}=s)\propto \exp(-L^d s^2/2\chi)$. Obviously, this result cannot hold at criticality because $\chi$ diverges and it is our aim to devise a formalism allowing us to compute this PDF perturbatively in the critical limit.

Since the PDF is a universal quantity, we choose to work in the continuum with the $\phi^4$ theory and with periodic boundary conditions. In the following, we present the formalism for the Ising case, the generalization to O($N$) being straightforward. The object of interest is therefore
\begin{equation}\label{def-pdf}
    P(\hat{s}=s) \propto\int D\hat{\phi}~\delta(\hat{s}-s)\exp\left(-\int_V\mathcal{H}[\hat{\phi}]\right),
\end{equation}
where 
\begin{equation}
    \hat s=\frac{1}{L^d}\int_V \hat{\phi}(x),
\end{equation}
with $\int_V=\int d^dx$ and the integral is limited to the volume $V=L^d$. The Hamiltonian is 
\begin{equation}\label{hamiltonian}
    \mathcal{H}\left(\hat{\phi}(x)\right)= \frac{1}{2}\left(\nabla\hat{\phi}(x)\right)^2+\frac{1}{2}Z_{2}\,t\,{\hat{\phi}}^2(x)+\frac{1}{4!}Z_{4}\,g\,{\hat{\phi}}^4(x).
\end{equation}
The bare couplings, that is, the couplings defined at the ultra-violet scale $\Lambda$, e.g. the inverse lattice spacing $\Lambda\sim a^{-1}$, are $t_0=Z_{2}\,t$ and $g_0=Z_{4}\,g$ while $t$ and $g$ are dimensionful renormalized couplings defined at a scale $\mu$ which is unspecified for the moment. In the following, we use dimensional regularization and the Minimal Subtraction scheme (MS) and  $Z_{2}$ and $Z_{4}$ have been introduced in the previous equation to cancel the divergences coming out of the loop-expansion when it is performed at fixed $t_0$ and $g_0$.\footnote{Notice that in MS, the bare critical mass is vanishing so that criticality is reached when $t_0\to0$, that is, $t\propto T-T_c$.} Dimensionless renormalized quantities, to be defined later in Eq.  \eqref{dimensionless}, will be denoted with a bar and for $\mu=L^{-1}$ by a tilde, see Eq. \eqref{stilde}. [Notice that the normalization factor in front of the PDF in Eq.~\eqref{def-pdf} has been omitted for simplicity.]

Replacing the Dirac delta function in Eq.~\eqref{def-pdf} by an infinitely peaked Gaussian, i.e. $\delta(z)\propto \exp(\frac{-M^2z^2}{2})$ with $M\rightarrow\infty$, the PDF can be interpreted as the partition function ($\mathcal{N}$ a normalization constant)
 \begin{equation}\label{b}
     \mathcal{Z}_{M,s}[h] =\mathcal{N}\int D\hat{\phi}\exp\left(-\int_V \mathcal{H}_{M,s}+\int_V h(x)\hat{\phi}(x)\right)
 \end{equation}  
 at vanishing magnetic field $h=0$  of a system with  Hamiltonian
 \begin{equation}
     \mathcal{H}_{M,s}\left(\hat{\phi}(x)\right)=\mathcal{H}\left(\hat{\phi}(x)\right)+ \frac{M^2}{2}\left(\int_{V}(\hat{\phi}(x)-s)\right)^2.
 \end{equation} 

Following \cite{Balog2022}, we now show that $P(s)$ can be conveniently written in terms of the modified Legendre transform defined by
\begin{equation}\label{legendre}
    \Gamma_{M}[\phi]+\log{\mathcal{Z}_{M,s}[h]} = \int_V h(x){\phi}(x)-\frac{M^2}{2}\left(\int_V({\phi}(x)-s)\right)^2,
\end{equation}
with
\begin{equation}
     \phi(x) = \frac{\delta \log{\mathcal{Z}_{M,s}[h]}}{\delta h(x)} ,
 \end{equation}
 and
\begin{equation}
   \frac{\delta \Gamma_{M}}{\delta\phi(x)} = h(x)-M^2\int_{y} (\phi(y)-s) .
\end{equation}
Using Eq.~\eqref{legendre}, we find
\begin{equation}\label{gammam}
     e^{-\Gamma_{M}[\phi(x)]} =\mathcal{N}\int  D\hat{\phi}~e^{-\int_V \mathcal{H}(\hat\phi(x))+\int_V \frac{\delta \Gamma_{M}}{\delta\phi}\cdot({\hat\phi}-\phi)-\frac{M^2}{2}\left(\int_V(\hat{\phi}-\phi)\right)^2}.
 \end{equation}  
Note that although it is not explicit from its definition,
Eq.~\eqref{legendre}, $\Gamma_{M}$ is independent of $s$ as can be checked
from Eq.~\eqref{gammam} or by deriving Eq.~\eqref{legendre} with respect to $s$.

The PDF $P(s)$ can be recovered from Eq.~\eqref{gammam} in the limit  $M\to\infty$
\begin{align}    
    \lim_{M\rightarrow\infty} e^{-\Gamma_{M}[\phi(x)=s]} &= \mathcal{N}\int D\hat{\phi}~e^{-\int \mathcal{H}[\hat{\phi}]+h.\int (\hat{\phi}-s)}\delta\left(\int{\hat\phi(x)}-s \right),\nonumber\\
    &\propto P(\hat{s}=s).
\end{align}
The rate function $I(s)$ is defined by
\begin{equation}
    P(\hat{s}=s) \propto e^{-L^d I(s,\,\xi_\infty,\,L)}.
\end{equation}
and thus 
\begin{equation}\label{gammam-rate}
   \lim_{M\to\infty}\Gamma_{M}[\phi(x)=s] =L^d I(s),
\end{equation}
where the parameters on which $\Gamma_{M}$ depends have not been specified, see below.

\section{The one-loop calculation \label{sec:1loop}}

The one-loop calculation of $I(s)$ requires to expand $\mathcal{Z}_{M,s}[h]$ at leading order around the mean-field configuration $\hat{\phi}_{0}$ defined by
\begin{equation}
    \frac{\delta }{\delta\hat{\phi}}\int_{V}\mathcal{H}_{M,s}\bigg|_{\hat{\phi}=\hat{\phi}_{0}}=h,
\end{equation}
and
\begin{equation}\label{one-loop}
     \mathcal{Z}_{M,s}[h]  = \mathcal{N}\exp\biggl(-\int_V \mathcal{H}_{M,s}\big(\hat{\phi}_{0}\big)+\int_V h.\hat{\phi}_{0}-\frac{1}{2} \Tr{\log\mathcal{H}^{(2)}_{M,s}+\dots\biggr)}.
\end{equation}
 
Using Eqs.~(\ref{legendre}) and (\ref{one-loop}), we find at one-loop
  \begin{equation}\label{gammam-one-loop}
      \Gamma_{M}[\phi] = \int_V \mathcal{H}(\phi)+\frac{1}{2} \Tr\log\mathcal{H}^{(2)}_{M,s}[\phi]-\log(\mathcal{N}).
  \end{equation}
The equation above is {\it a priori} ill-defined because the ${\Tr\log}$ term is divergent. However, once it is dimensionally regularized and the $\log(\mathcal{N})$ term is chosen such that it is equal to the value of $\frac{1}{2}\Tr\log\mathcal{H}^{(2)}_{M,s}[\phi]$ computed at vanishing field and vanishing mass (i.e. at $t\propto T-T_{c}=0$) with an addition of a constant counter-term, it becomes well-defined. 

We thus find (up to a constant divergent term)
\begin{equation}\label{logN}
   \frac{1}{2}\Tr\log\mathcal{H}^{(2)}_{M,s}[\phi]-\log(\mathcal{N})=\frac{1}{2 L^{d}}\sum_{q}\log\left(1+\frac{t+g{\phi}^2/2}{q^{2}+M^{2}L^{d}\delta_{q, 0}}\right) ,
\end{equation}
where $q$ is a $d-$dimensional vector with components $q_i=2\pi n_i/L$ and $n_i\in {\mathbb Z}$. We thus obtain
\begin{align}\label{e}
     \lim_{M\rightarrow\infty}&\Gamma_M[\phi(x)=s]  = L^d\biggl(\frac{1}{2}Z_{2}{t} s^2+\frac{1}{4!}Z_{4}{g} s^4\nonumber\\
     & +\frac{1}{2 L^d}\sum_{q\neq 0}\log\left(1+\frac{{t}+{g} s^2/2}{q^2}\right)+\frac{{t}^{\,2}L^{\epsilon}}{32{\pi^2\epsilon}}\biggr),
\end{align}
where the last term in Eq.~\eqref{e} is precisely the constant counter-term mentioned before Eq.~\eqref{logN}. The renormalized coupling ${g}$ a priori depends on a scale $\mu$. 
Since there cannot be any fluctuation with wavelengths larger than the system size, it is natural to choose the scale $\mu=L^{-1}$ so as to include all fluctuations between the UV scale $\Lambda$ and the infrared scale $L^{-1}$. This is implicit in Eq.~\eqref{e}. 

Using the dimensionless variables
\begin{equation}\label{stilde}
    \tilde{s}=sL^{\frac{d-2+\eta}{2}}\ , \ \tilde{g}=L^{4-d}{g}(\mu=L^{-1}),
\end{equation}
we obtain at first order in $\epsilon=4-d$, putting the anomalous dimension $\eta=0$ at this order
   \begin{align}\label{m}
  \lim_{M\rightarrow\infty}\Gamma_M[\tilde{s},&\,  \mu=L^{-1}]  = \frac{1}{2}Z_{2}({t}L^{2})\tilde{s}^{\,2}+\frac{1}{4!}Z_{4}\tilde{g}\tilde{s}^{\,4}+\frac{(tL^{2})^{2}}{32{\pi^2\epsilon}}\nonumber\\
  & +\frac{1}{2}\sum_{n\neq 0}\log\left(1+\frac{2(tL^{2})+\tilde{g}\tilde{s}^{\,2}}{8\pi^2n^2}\right),
  \end{align}
  where $n\in{\mathbb Z}^d$.
 In dimension ${d=4-\epsilon}$ (see Appendix \ref{app1} and \cite{Rudnick1985,Eisenriegler1987}),
 \begin{equation}\label{f}
 \sum_{n\neq 0} \log\left(1+\frac{x}{n^2}\right)+(x\pi)^{d/2}\Gamma(-d/2)= \Delta(x)+O(\epsilon),
 \end{equation}
 with $$\Delta(x)=\theta(x \pi )-\theta(0),$$
 and 
 \begin{align}
\theta(z)=-\int_0^\infty d\sigma \frac{e^{-\sigma z}}\sigma\left(\vartheta^d(\sigma)-1-\sigma^{-d/2}\right),
 \end{align}
where $\vartheta(x)$ is the Jacobi $\vartheta$ function.\\
 
By definition, $Z_{2}$ and $Z_{4}$ are series in the dimensionless coupling constant $\tilde{g}$ designed in the minimal scheme to cancel out the poles coming out of the term $(x\pi)^{d/2}\Gamma(-d/2)$. Thus, we have
 \begin{align}
 \lim_{M\rightarrow\infty}\Gamma_M&[\tilde{s}, L^{-1}]  = \frac{1}{2}({t}L^{2})\tilde{s}^{\,2}+\frac{2\pi^2 }{3}\tilde{u}\tilde{s}^{\,4}\nonumber\\
 &+\frac{\pi^{2}}{4}\biggl(\frac{{t}L^{2}}{4\pi^2} +2\tilde{u}\tilde{s}^{\,2}\biggr)^2\left[\gamma+\log2\pi-\frac{3}{2}\right.\nonumber\\
  &+\left.\log\biggl(\frac{{t}L^{2}}{8\pi^2}+\tilde{u}\tilde{s}^{\,2}\biggr)\right]+\frac{1}{2}\Delta\left(\frac{{t}L^{2}}{4\pi^2}+2\tilde{u}\tilde{s}^{\,2}\right) + O(\epsilon),
 \end{align}
 where $\gamma$ is the Euler-Mascheroni constant,  $\tilde{g}={16\pi^2}\tilde{u}$ and the counter terms $Z_{2}$ and $Z_{4}$ are found to be
\begin{equation}\label{Countert}
     Z_{2}=1+\frac{\tilde{g}}{16\pi ^2\epsilon}+O(\tilde{g}^2)\ , \ \ Z_{4}=1+\frac{3\tilde{g}}{16\pi ^2\epsilon}+O(\tilde{g}^2).
\end{equation}
 Universality takes place in the double limit of infinite volume, $L\to\infty$, and infinite bulk correlation length, $\xi_\infty\to\infty$, that behaves as $\xi_\infty\sim {t}^{-1/\nu}$. We show below that this limit is not unique  and that keeping the ratio $\zeta=\lim_{L,\xi_\infty\rightarrow\infty}L/\xi_{\infty}$ fixed, there are infinitely 
 many inequivalent ways of taking this double limit and thus infinitely many universal rate functions indexed by $\zeta$ 
 \begin{equation}\label{def-rate-scaling}
    I_{\zeta}(\tilde{s})=\lim_{M,L,t^{-1}\rightarrow\infty}\Gamma_M[\tilde{s}, L^{-1}] \quad {\rm at\  fixed} \quad \zeta=\lim_{L,\xi_\infty\rightarrow\infty}\frac{L}{\xi_{\infty}} . 
 \end{equation}

 The derivation of $I_{\zeta}(\tilde{s})$ proceeds in two steps. Firstly, if we were running the RG flow, we would obviously find that the dimensionless coupling constant $\tilde{u}$ flows to its fixed point value $u_{*}=\frac{3\epsilon}{N+8}$ (with $N=1$ for Ising model) when $\mu=L^{-1}\to 0$. Secondly, since for $\mu\ll\Lambda$,  $\xi_\infty^{-1}=\mu\left(t/{\mu ^{2}}\right)^{\nu}$ and we have chosen $\mu=L^{-1}$, we find that ${t}L^{2}=\zeta^{1/\nu}$. Here and in the following, in all one-loop expressions we use the one-loop exponents, e.g. for $\nu$, $1/\nu=2-\epsilon/3$ and $\eta=0$. Thus, by defining $x=\sqrt{u_{*}}\,\tilde{s}$, we  finally find
  \begin{align}\label{a3}
 I_{\zeta}(x) & =  \frac{3}{\epsilon}\left(\frac{1}{2}\zeta^{\frac{1}{\nu}} x^2+\frac{2\pi^2 }{3}x^4\right)\nonumber\\
 & +\frac{\pi^{2}}{4}\left(\frac{\zeta^{\frac{1}{\nu}}}{4\pi^2}+2x^2\right)^2\biggl(\gamma+\log2\pi-\frac{3}{2} +\log\biggl(\frac{\zeta^{\frac{1}{\nu}}}{8\pi^2}+x^2\biggr)\biggr)\nonumber\\
  &+\frac{1}{2}\Delta\left(\frac{\zeta^{\frac{1}{\nu}}}{4\pi^2}+2x^2\right) + O(\epsilon),
 \end{align}
 which shows that the universal rate functions are only functions of $\zeta$ and not of $L$ and $\xi_\infty$ separately. Notice that using the variable $x$ instead of $\tilde s$ is on one hand a finite redefinition of the field when $\epsilon$ is fixed and nonvanishing and on the other hand is what guarantees that Eq.~\eqref{a3} is the beginning of a systematic $\epsilon$-expansion \cite{PhysRevB.7.232, PhysRevLett.29.591}. 
 
 The above result can be easily extended to the $O(N)$ case 
 \begin{align}\label{a4}
      I_{\zeta, N}&(x)  =  \frac{N+8}{3\epsilon}\left(\frac{1}{2}\zeta^{\frac{1}{\nu}} x^2+\frac{2\pi^2 }{3}x^4\right)+\frac{1}{2}\Delta\left(\frac{\zeta^{\frac{1}{\nu}}}{4\pi^2}+2x^2\right)\nonumber\\
 & +\frac{\pi^{2}}{4}\left(\frac{\zeta^{\frac{1}{\nu}}}{4\pi^2}+2x^2\right)^2\biggl(\gamma+\log2\pi-\frac{3}{2} +\log\biggl(\frac{\zeta^{\frac{1}{\nu}}}{8\pi^2}+x^2\biggr)\biggr)\nonumber\\
  &+(N-1)\biggl[\frac{\pi^{2}}{4}\left(\frac{\zeta^{\frac{1}{\nu}}}{4\pi^2}+\frac{2x^2}{3}\right)^2\biggl(\gamma+\log2\pi-\frac{3}{2} \nonumber\\
  &+\log\biggl(\frac{\zeta^{\frac{1}{\nu}}}{8\pi^2}+\frac{x^2}{3}\biggr)\biggr)+\frac{1}{2}\Delta\left(\frac{\zeta^{\frac{1}{\nu}}}{4\pi^2}+\frac{2x^2}{3}\right)\biggr]+ O(\epsilon),
 \end{align}

The calculation of the rate functions \eqref{a3} and \eqref{a4} has the advantage of simplicity and is based on our intuition that choosing $L^{-1}$ as the renormalization scale is the best we can do. It has the disadvantage of hiding the fact that these functions should be scaling functions which is not fully clear in \eqref{a3} and \eqref{a4}. Moreover, we shall see in  Section \ref{MC} that the large field behavior of the one-loop rate function is qualitatively incorrect, see Fig.~\ref{true-zeta=0} for the special case of $\zeta=0$. The reason is clear: at large field, the rate function behaves as the fixed point potential, that is, as $\tilde s^{\,\delta+1}$ with $\delta=\frac{d+2-\eta}{d-2+\eta}$, \cite{Balog2022} whereas perturbation theory yields the same result but fully expanded in $\epsilon$, that is, both $\delta$  and $\tilde{s}^{\delta+1}$ are expanded. This explains the presence of logarithms, visible in Eq.~\eqref{a3} and \eqref{a4}, that become large at large fields and that invalidate perturbation theory in this range of fields. They are an internal signal of perturbation theory that it fails at large fields. The recipe for overcoming this problem is well known: these logarithms must be resummed to recreate the power law $\tilde s^{\,\delta+1}$ where  $\delta$ takes its one-loop value. This is automatic in the NPRG formalism, but not in perturbation theory. In what follows, we show how to improve our determination of $I_\zeta(\tilde s)$ given in Eq.~\eqref{a3} at large values of the field.

 \section{Derivation of Scaling Function: An Alternative Route \label{sec:alternative}}

Before looking at possible improvements to the rate functions at large fields, we must first show that they are scaling functions and determine them. To do this, we generalize the calculation of the previous section by keeping $\mu$ arbitrary, which will enable us to put these functions into a scaling form.
 We will then show that this is not sufficient to recover the correct large-field behavior and, in a second step, we will present various improvement schemes that will enable us to recover the correct large-field power-law behavior. 

The fact that the rate function $I_\zeta(s)$ is a scaling function is a consequence of Eq.~\eqref{gammam-rate} that shows that it can be seen as the free energy of the model with Hamiltonian $ \mathcal{H}_{M,s}$ in the limit of large $M$. Defining $\delta\Gamma_M(s,T,L)=\Gamma_M(s,T,L)-\Gamma_M(0,T_c,L)$, its singular part is  universal and its density $\delta\gamma_M$ is proportional to
\begin{equation}
   \frac{\delta\Gamma_M(s,T,L)}{(L\,\mu)^d}=\delta\gamma_M(s,T,g,L),
\end{equation}
with $\delta\gamma_M(s,T,g,L)$ a dimensionless function. This function can be written in terms of dimensionless renormalized parameters defined at a scale $\mu$ by
\begin{equation} \label{dimensionless}
\begin{split}
\bar{g}=&\mu^{-\epsilon}g =\mu^{-\epsilon}16\pi^2 \bar{u},\\
 \bar{s}=&\mu^{-1+\epsilon/2}Z^{-1/2}s,\\
\bar{t}=&\mu^{-2}t,\\
\bar{L}=&\mu L,
\end{split}
\end{equation}
where $Z=Z(\bar{u})$ is the field renormalization. Taking first the limit $M\to\infty$, it satisfies
\begin{equation}\label{relation-scaling}
\mu^d \delta\gamma_\infty(\bar{s},\bar{t},\bar{g},\bar{L})=   {\mu'}^d \delta\gamma_\infty(\bar{s}',\bar{t}\,',\bar{g}',\bar{L}'),
\end{equation}
where the prime variables are defined at scale $\mu'$. Defining $l=\mu'/\mu$, the evolution of the couplings and fields with $l$ is given by their RG flow. Provided we take the thermodynamic limit ($\bar{L}\to\infty$) as well as the critical limit ($\bar{t}\to0$) and we choose $l$ small, that is, we consider large length scales compared to the UV cutoff, the coupling $\bar{u}$ runs to its fixed point value $u^*$ and $Z(u^*)\sim \mu^{-\eta}$ where $\eta$ is the anomalous dimension. Choosing $l=\bar{L}^{-1}$, we obtain from Eqs.~\eqref{dimensionless} and~\eqref{relation-scaling}
\begin{equation}\label{scaling-function}
\begin{split}
 \delta\gamma_\infty(\bar{s},\bar{t},\bar{u},\bar{L})&=l^d\delta\gamma_\infty(\bar{s}\,l^{-\beta/\nu},\bar{t}\,l^{-1/\nu},u^*, \bar{L}\,l),\\
 &= \bar{L}^{-d}\delta\gamma_\infty(\tilde{s}, \zeta^{1/\nu},u^*,1),
\end{split}
\end{equation}
where $\tilde{s}$ is defined in Eq.~\eqref{stilde} and $\beta/\nu=(d-2+\eta)/2$. Equation~\eqref{scaling-function} shows that the rate function is a scaling function of the two variables $\tilde{s}$ and $\zeta^{1/\nu}$.
 
To make explicit the fact that $I_\zeta(\tilde s)$ is a scaling function, we repeat our one-loop result for $\Gamma_M[\phi(x)=s]$ but this time with couplings defined at an arbitrary scale $\mu$ instead of $\mu=L^{-1}$, as we did in the previous section.

At this order, $\Gamma_{M}$ becomes
\begin{align}
\lim_{M\rightarrow\infty}&\Gamma_M[\phi(x) =\mu^{1-\epsilon/2}\bar{s}] = \bar{L}^d\biggl(\frac{1}{2}Z_{2}\bar{t}\bar{s}^2+\frac{2\pi^2}{3}Z_{4}\bar{u}\bar{s}^{\,4}\nonumber\\
& +\frac{1}{2\bar{L}^d}\sum_{n\neq 0}\log\biggl(1+\frac{\frac{\bar{t}\bar{L}^2}{4\pi^2}+2\bar{u}\bar{s}^2\bar{L}^2}{n^2}\biggr)+\frac{\bar{t}^{\,2}\bar{L}^{\epsilon}}{32{\pi^2\epsilon}}\biggr),
\end{align}
where once again $Z_{2}$ and $Z_{4}$ are counterterms expanded in powers of $\bar{u}$ to cancel out the poles  coming out of the one-loop term and   $n\in{\mathbb Z}^d$. 
One can then once again use Eq.~(\ref{f}), to compute the discrete sum in the above expression in $4-\epsilon$ dimensions, and thus end up with
\begin{align}\label{g}
\lim_{M\rightarrow\infty}&\bar{L}^{\,-d}\Gamma_M[\phi(x) =\mu^{1-\epsilon/2}\bar{s}]  =\nonumber\\
&\biggl(\frac{1}{2}\bar{t} \bar{s}^{\,2}+\frac{2\pi^2}{3}\bar{u}\bar{s}^{\,4}+\frac{\pi^{2}}{4}\biggl(\frac{\bar{t}}{4\pi^2}+2\bar{u}\bar{s}^{\,2} \biggr)^2\biggl(\gamma+\log2\pi\nonumber\\
&-\frac{3}{2}+\log\biggl(\frac{\bar{t}}{8\pi^2}+\bar{u}\bar{s}^{\,2}\biggr)\biggr)\biggr)\nonumber\\
&+ \frac{1}{2\tilde{L}^{\,4}}\Delta\left(\left[\frac{\bar{t}}{4\pi^2}+2\bar{u} \bar{s}^{\,2}\right]\bar{L}^{\,2}\right).
\end{align}
Now defining the variable  $\bar{x}=\sqrt{\bar{u}} \bar{s}$ and  taking the  limit of criticality and infinite volume together by keeping the ratio $\zeta=\lim_{L,\xi_\infty\rightarrow\infty}L/\xi_{\infty}=\lim_{\bar{t}\rightarrow 0, \bar{L}\rightarrow\infty}\bar{t}^{\,\nu}\bar{L}$ fixed and sending $\bar{u}$ to its fixed point value $u_*=\frac{3\epsilon}{n+8}$, with $n=1$, we get at one loop, that is, up to terms of order $\epsilon$ 
 \begin{align}\label{o}
&\lim_{M\rightarrow\infty}\bar{L}^{\,-d}\Gamma_M(\bar{x}, \zeta,\bar{L})
 = \nonumber\\
&\ \frac{3}{\epsilon}\biggl[\frac{1}{2}\biggl(\frac{\zeta}{{\bar{L}}}\biggr)^{\frac{1}{\nu}}{\bar{x}^2}+\frac{2\pi^2}{3}{\bar{x}^4}\biggr] + \frac{\pi^2}{4}\biggl(\frac{1}{4\pi^2}\biggl(\frac{\zeta}{{\bar{L}}}\biggr)^{1/\nu}+{2\bar{x}^2}\biggr)^2\biggl[\gamma+\log2\pi\nonumber\\
&\ -\frac{3}{2}+\log\left(\frac{\bigl(\zeta/\bar{L}\bigr)^{\frac{1}{\nu}}}{8\pi^2}+{\bar{x}^2}\right)\biggr] + \frac{1}{2\bar{L}^{\,4}}\Delta\left(\left[\frac{\bigl(\zeta/\bar{L}\bigr)^{\frac{1}{\nu}}}{4\pi^2}+2\bar{x}^2\right]\bar{L}^{\,2}\right) .
\end{align}
From the discussion above, we know that the right scaling variable for the rate function is of the form   $\tilde{s}=L^{\beta/\nu}s=L^{\frac{d-2+\eta}{2}}s$, see Eq.~\eqref{stilde}, or equivalently   
\begin{equation}\label{def-tildex}
 \tilde{x}=\bar{L}^{\frac{\beta}{\nu}}\bar{x}. 
\end{equation}
From now on, for notational ease, we trade $\tilde{x}$ for $x$. Thus, what we want to actually create is a scaling function of the variables $\zeta^{1/\nu}$ and $x$ which has a systematic $\epsilon$-expansion. For any function $f$  of scaling variables $x$ and $\zeta$, one finds at order $\epsilon$
\begin{align}\nonumber
f(\zeta^{\frac{1}{\nu}}, x) =&f(\zeta^{2-\frac{\epsilon}{3}}, \bar{L}^{1-\frac{\epsilon}{2}}\bar{x}),\\
 =& f(\zeta^{2}, \bar{x}\bar{L})-{\zeta}^{2}\frac{\epsilon}{3}\log{\zeta}\frac{\partial f}{\partial \zeta^{2}}|_{\zeta^{2}, \bar{x}\bar{L}}\nonumber\\
&-\epsilon\frac{\bar{x}\bar{L}}{2}\log{\bar{L}}\frac{\partial f}{\partial (\bar{x}\bar{L})}|_{\zeta^{2}, \bar{x}\bar{L}}+O(\epsilon^2),\label{q}
\end{align} 
Now, since the scaling function $f$ itself has an $\epsilon$-expansion, i.e. one can write $f$ as
\begin{equation*}
f(x)=\frac{f_{0}(x)}{\epsilon}+f_{1}(x)+O(\epsilon).
\end{equation*}
Thus, Eq.~(\ref{q}), can be written as
\begin{align}\label{scaling-expanded}
f(\zeta^{\frac{1}{\nu}}, x)& = \frac{f_{0}(\zeta^{2}, \bar{x}\bar{L})}{\epsilon}+f_{1}(\zeta^2, \bar{x}\bar{L})-\frac{{\zeta}^{2}}{3}\log{\zeta}\frac{\partial f_{0}}{\partial \zeta^{2}}|_{\zeta^{2}, \bar{x}\bar{L}}\nonumber\\
&-\frac{\bar{x}\bar{L}}{2}\log{\bar{L}}\frac{\partial f_{0}}{\partial (\bar{x}\bar{L})}|_{\zeta^{2}, \bar{x}\bar{L}}+O(\epsilon),
\end{align}
Using Eqs.~\eqref{def-rate-scaling},~\eqref{o} and ~\eqref{scaling-expanded}, we find
\begin{equation}
f_{0}(\zeta^{\frac{1}{\nu}}, x)=\frac{3}{2}\zeta^{\frac{1}{\nu}}{x}^2+2\pi^2 x^4,
\end{equation}
and 
\begin{align}
f_{1}(\zeta^{\frac{1}{\nu}}, x) & = \frac{\pi^2}{4}\left(\frac{\zeta^{\frac{1}{\nu}}}{4\pi^2}+2x^2\right)^2\left[\gamma+\log{2\pi}-\frac{3}{2}+\log{\left(\frac{\zeta^{\frac{1}{\nu}}}{8\pi^2}+x^2\right)}\right]\nonumber\\
& +\frac{1}{2}\Delta\left(\frac{\zeta^{\frac{1}{\nu}}}{4\pi^2}+2x^2\right).
\end{align}
Thus, one can write the total rate function $I_{\zeta}(x)$ as \footnote{\label{f1} There is a little subtlety here. Actually, there is an extra term of the form $\zeta^4\log{\bar{L}}$ in the expression below which cannot be recast in the scaling form. However, if one considers the perturbative expansion of the quantity $f(\zeta^{\frac{1}{\nu}}, x)-f(\zeta^{\frac{1}{\nu}}, 0)$, this term no longer appears and the identification of $f_0$ and $f_1$  works fine. This does not cause any problem since this constant term can be easily absorbed in the normalization constant of the PDF and we can in principle just work with the perturbative expansion of $f(\zeta^{\frac{1}{\nu}}, x)-f(\zeta^{\frac{1}{\nu}}, 0)$ instead of just $f(\zeta^{\frac{1}{\nu}}, x)$. This is reminiscent of the choice of the $\log{\mathcal{N}}$ term in Eq. \eqref{e}.}
\begin{align}\label{rate-scaling}
I_{\zeta}(x) & =  \frac{3}{\epsilon}\left(\frac{1}{2}\zeta^{\frac{1}{\nu}} x^2+\frac{2\pi^2 }{3}x^4\right)+\frac{\pi^{2}}{4}\left(\frac{\zeta^{\frac{1}{\nu}}}{4\pi^2}+2x^2\right)^2\biggl(\gamma+\log2\pi\nonumber\\
& -\frac{3}{2}+\log\left(\frac{\zeta^{\frac{1}{\nu}}}{8\pi^2}+x^2\right)\biggr)+\frac{1}{2}\Delta\left(\frac{\zeta^{\frac{1}{\nu}}}{4\pi^2}+2x^2\right) + O(\epsilon),
\end{align}
An important remark is in order here: Eq.~(\ref{rate-scaling}) is exactly the same as Eq.~(\ref{a3}). We have checked that this also holds true in  the $O(N)$ case and that Eq.~$(\ref{a4})$ corresponds to a scaling function expanded in $\epsilon$. This means that our choice of scale $\mu=L^{-1}$, performed in Eq.~\eqref{m} on physical ground to optimize the determination of the rate function, is in fact what makes it a scaling function.  It also means that recasting $I_\zeta$ under the form of a scaling function is not sufficient to get rid of the large logarithms appearing at large fields. 

\section{The large field behavior and the RG improvements}\label{improvement}

To analyze the field dependence of the rate function, it is more convenient to work with $I_{\zeta}(x)-I_{\zeta}(0)$ instead of  $I_{\zeta}(x)$, since the constant $I_{\zeta}(0)$ only changes the PDF's normalization (see also footnote \ref{f1}). Using Eq.~(\ref{a3}), we find that at one loop, this quantity is of the form
 \begin{equation}
  I_{\zeta}(x)-I_{\zeta}(0)=\frac{S_{0}(x)}{\epsilon}+S_{1}(x)+O(\epsilon),
 \end{equation}
with $S_i(x)=f_i(x)-f_i(0)$ and $f_i$ introduced in the previous section. 
  
This perturbation expansion is only valid if $S_{1}(x)/S_{0}(x)$ is finite for all values of $x$. However, this quantity diverges ${S_{1}(x)}/{S_{0}(x)}$ at large field which invalidates perturbation theory in this domain. This calls for an improvement and the aim of this section is to cure the divergence that appears in the large field behaviour of $I_{\zeta}(x)$ using RG. 
  
To do so, we need to take one step back. Our starting point is Eq.~(\ref{g}). Near criticality where $u$ can be replaced by its fixed point value $u_*$, it reads 
\begin{align}\label{gamma-one-loop-y-tbar}
   \lim_{M\rightarrow\infty}&{\bar{L}}^{-d}\Gamma_M(\bar{x}, \bar{t},\bar{L})
= \frac{3}{\epsilon}\biggl[\frac{1}{2}\bar{t}{\bar{x}^2}+\frac{2\pi^2}{3}{\bar{x}^4}\biggr]\nonumber\\
& + \frac{\pi^{2}}{4}\biggl(\frac{1}{4\pi^2}\bar{t}+{2\bar{x}^2}\biggr)^2\biggl(\gamma+\log2\pi-\frac{3}{2}\nonumber\\
&+\log\left(\frac{\bar{t}}{8\pi^2}+{{\bar{x}}^2}\right)\biggr)+ \frac{1}{2\bar{L}^4}\Delta\left(\left[\frac{\bar{t}}{4\pi^2}+2\bar{x}^2\right]\bar{L}^2\right)\nonumber\\  
&  + O(\epsilon)  ,
\end{align}
 with $\bar{x}\simeq \sqrt{u_{*}}\bar{s}$.

As explained above, we compute from this expression the ratio $S_{1}(\bar{x}, \zeta,\bar{L})/S_{0}(\bar{x}, \zeta,\bar{L})$ in the limit $\bar{t}\rightarrow 0$ and $\bar{L}\rightarrow \infty$ and study when it becomes large. Let us notice that, as already said above, instead of $\Gamma_{M}(\bar{x}, \bar{t}, \bar{L})$, it is more convenient to consider $\Gamma_{M}(\bar{x}, \bar{t}, \bar{L})-\Gamma_{M}(0, \bar{t},\bar{L})$.

\begin{table}[t]
  \begin{center}
    \begin{tabular}{ c|c } 
      \text{Limiting Conditions} 
       &$S_1/S_0$\\
      \hline
       \\
     $(i)~~\bar{t}\bar{L}^2\ll 8\pi^2 \bar{x}^2\bar{L}^2\ {\rm and}\  \bar{x}^2\bar{L}^2\ll1$&$\log{\bar{L}^{-2}}$ \\
     &\\
     $(ii)~~\bar{t}\bar{L}^2\ll 8\pi^2 \bar{x}^2\bar{L}^2\ {\rm and}\ \bar{x}^2\bar{L}^2\gg1$&$\log{\bar{x}^{2}}$
    \end{tabular}
    \caption{ \label{table-case} Behavior of the ratio $S_1/S_0$.}
  \end{center}
\end{table}
We find that two different ranges of parameters are problematic for the ratio $S_1/S_0$ in the double limit  $\bar{t}\to 0$ and $\bar{L}\to\infty$. They are summarized in Table \ref{table-case}, where we show that since we are working in the thermodynamic limit, the condition $\bar{x}^2{\bar{L}}^2\gg1$ corresponds to the large field limit while that of  $\bar{x}^2{\bar{L}}^2\ll1$ corresponds to the small field region, i.e. $\bar{x}\rightarrow 0$. We see in Table \ref{table-case} that in both cases the perturbation expansion is riddled with logarithmic divergences. The present section aims to show how to get rid of them.

We again use
\begin{equation}
\begin{split}
 \bar{s}'&=\bar{s}\, l^{-\beta/\nu}, \\   
  \bar{t}\,'&=\bar{t}\,l^{-1/\nu}, \\   
   \bar{L}'&=\bar{L}\,l .
\end{split}
\end{equation}
Then, choosing $l$ small enough, Eq.~\eqref{gamma-one-loop-y-tbar} can be written at criticality
\begin{align}\label{a6}
\lim_{M\rightarrow\infty}\Gamma_M[\bar{x}{}', \bar{t}\,', \bar{L}'] & ={\bar{L}'{}}^d\biggl(\frac{3}{\epsilon}\biggl[\frac{1}{2}\bar{t}\,'\bar{x}{}'^{2}+\frac{2\pi^2}{3}\bar{x}{}'^{4}\biggr]\nonumber\\
& +\frac{\pi^{2}}{4}\biggl(\frac{\bar{t}\,'}{4\pi^2} +2 \bar{x}{}'^{2}\biggr)^2\biggl(\gamma+\log2\pi\nonumber\\
&-\frac{3}{2}+\log\biggl(\frac{\bar{t}\,'}{8\pi^2}+\bar{x}{}'^{2}\biggr)\biggr)\nonumber\\
&+ \frac{1}{2{\bar{L}'{}}^4} \Delta\left(\left[\frac{\bar{t}\,'}{4\pi^2}+2\bar{x}{}'^2\right]{\bar{L}'{}}^2\right)\biggr),
\end{align}
with $\bar{x}{}'=\sqrt{u_{*}}\bar{s}'$, just as in the previous section, where once again we have a systematic $\epsilon$-expansion.  We now fix $\bar{L}'$ to get rid of the divergences that appear in both the large-field and small-field limits. 

One crucial thing that we have already noticed, is that when we set $\bar{L}'=1$, that is, when the RG flow is run down to the momentum scale $\mu'=L^{-1}$, Eq.~(\ref{a6}) reverts back to Eq.~(\ref{a3}) where there is no longer any problem at small field. However, the problem survives at large field because  the $\tilde{s}^{\delta+1}$ behaviour is not reproduced. In principle, the presence of large logarithms in perturbative series satisfying scaling is cured by the RG, which allows the resummation of these logarithms, transforming the expansion of the quantity studied into an expansion of the exponent of its power law behavior. The difficulty here lies in the fact that the PDF is a function, not a number, and that this RG improvement must be functional. In other words, the $\bar{L}'$ scale must depend on the field $\tilde{s}$ in such a way that the $\log \bar{x}^2$ term shown in Table \ref{table-case} disappears. This means, in terms of the  variable $x$ defined in Eq.~\eqref{def-tildex}, the behaviour of $\bar{L}'$ at large value of $x$ should go as $x^{\frac{\nu}{\beta}}$. From this, we come to the conclusion that $\bar{L}'(x)$ should behave as $1$ in the small field region and as $x^{\frac{\nu}{\beta}}$ in the large field region.

Writing the final result in terms of the quantities that are kept fixed in the critical limit and in the thermodynamic limit, that is,
$x$ and $\zeta$, we get for the improved rate function
\begin{align}\label{IMPrate}
 I_{\zeta}(x) & =\frac{3{\bar{L}'{}}^d}{\epsilon}\left[\frac{1}{2}\left(\frac{\zeta}{\bar{L}'{}}\right)^{\frac{1}{\nu}}\frac{x^2}{\bar{L}'{}^{\frac{2\beta}{\nu}}}+\frac{2\pi^2}{3}\frac{x^4}{\bar{L}'{}^{\frac{4\beta}{\nu}}}\right]\nonumber\\
  &+\bar{L}'{}^{d}\biggl[\frac{\pi^{2}}{4}\left(\frac{1}{4\pi^2}\left(\frac{\zeta}{\bar{L}'{}}\right)^{1/\nu}+\frac{2x^2}{\bar{L}'{}^{\frac{2\beta}{\nu}}}\right)^2\biggl(\gamma+\log{2\pi}-\frac{3}{2}\nonumber\\
  & +\log\left(\frac{\left(\frac{\zeta}{\bar{L}'{}}\right)^{1/\nu}}{8\pi^2}+\frac{x^2}{{\bar{L}'{}^{\frac{2\beta}{\nu}}}}\right)\biggr)\biggr]\nonumber\\
 & +\bar{L}'{}^d\left[\frac{1}{2\bar{L}'{}^4} \Delta\left(\left[\frac{\left(\frac{\zeta}{\bar{L}'{}}\right)^{1/\nu}}{4\pi^2}+2\frac{x^2}{\bar{L}'{}^{\frac{2\beta}{\nu}}}\right]\bar{L}'{}^2\right)\right] + O(\epsilon),
\end{align}
where $\bar{L}'$ has the desirable small field and large field behavior as discussed above. 

One must notice that there is not just one but infinitely many ways to carry out the RG improvement depending upon the functional form of $\bar{L}'(x)$. We choose a family of RG improvements indexed by a parameter $\alpha$
\begin{equation}\label{IMPE}
\bar{L}'(x, \alpha)=1+\alpha x^{\nu/\beta}e^{-\frac{\zeta^{1/\nu}}{8\pi^{2}x^{2}}}    .
\end{equation}
Notice that at one loop and in $d=3$, $\nu/\beta=2$. Notice also that we have chosen a $\zeta$-dependent function $\bar{L}'$. The reasoning behind this choice is as follows. As we increase the value of $\zeta$, the rate at which the rate-function grows increases. We therefore need to design an improvement in such a way that it can smoothly interpolate between the large field and the small field behaviour. Due to this we must encode in our improvement some kind of damping with $\zeta$. As is evident in Eqs.~\eqref{IMPrate} and~\eqref{a3}, there always exists a competition between $\zeta^{1/\nu}/8\pi^{2}$ and $x^{2}$ and it is only after $x^{2}$ becomes much larger than $\zeta^{1/\nu}/8\pi^{2}$ that  the large field limit is truly reached. It is by keeping these factors in mind that the improvement in~\eqref{IMPE} has been designed.

Let us notice that the family of improvements parameterized by $\alpha$, Eq.~\eqref{IMPE}, differ quantitatively but they all yield the correct large field behavior of the rate function which is proportional to $\tilde{s}^{\delta+1}$, with $\delta+1=6$ (instead of 5.79), whereas the unimproved rate function behaves at large field as $\tilde{s}^4$ with large logarithmic corrections.

Before comparing with Monte Carlo simulations, we have to decide how to fix $\alpha$. We use the Principle of Minimal Sensitivity (PMS) that states that since the exact rate function should be independent of the RG improvements, the best we can do for fixing $\alpha$ is to choose the value where the dependence of $I_\zeta(x)$ on $\alpha$ is the smallest, that is, the value of $\alpha$ for which it is stationary. This principle has been used in the NPRG context and other contexts, see \cite{StevensonBook} and references therein, and it has made it possible to determine the values of critical exponents of the O($N$) models very accurately \cite{Balog2019,DePolsi2022}. For $I_\zeta(x)$ considered as a function of $\alpha$, the difficulty is that it is a function and not a number that has to be stationary and it is {\it a priori} not trivial that  $I_\zeta(x)$ is stationary in $\alpha$ for all values of $x$ simultaneously. However, this is what we find with high precision and it allows us to select a single optimal value of $\alpha$, see Section~\ref{improved-results}. 

The RG improvement also enables us to predict a non-trivial behavior of the tails of the PDF. Consider Eq.~\eqref{IMPrate}, one can see that in the large field limit i.e. $x\rightarrow\infty$, the leading behavior of $I_{\zeta}(x)$ is given by the power law $x^{\delta+1}$. This is what is expected since that is what we designed the RG improvement for. What is striking is that, in the large field regime, i.e. $x\to\infty$  there exists a sub-leading behavior of the rate function which goes as $-\log(x)=-{\frac{\delta-1}{2}}\log(x)+O(\epsilon)$. This means that in the large field regime, the PDF must go as $x^{\frac{\delta-1}{2}}e^{-x^{\delta+1}}$. This behavior has been previously predicted in the Ising model some time ago \cite{Hilfer1995,Bruce1995}, and is, in fact, generic for equilibrium systems \cite{next_paper_2035, Stella2023}.

\section{Analytic Continuation to the Low-Temperature Phase\label{sec:lowT}}
In the present section, we discuss how our results can be generalized to the approach to criticality from the low-temperature phase: $T\to T_c^-$. This corresponds to the case $t<0$, which we parametrize by a negative $\zeta$, i.e. we generalize it to $\zeta = {\rm sgn}(t)L/\xi_\infty(|t|)$.\footnote{Note that for $t<0$, $\xi_\infty(|t|)$ is not the correlation length of the system since it is infinite for $N>1$ and different from that of the disordered phase for $N=1$. It is however a convenient way to universally parameterize the family of rate functions in this phase.} As $t$ is traded for $\zeta^{1/\nu}$ on the disordered side, we  trade it for $-|\zeta|^{1/\nu}$ on the ordered side.
Although the entire family of rate functions for negative $\zeta$'s still remains inaccessible in the current approach, we show that not all is lost in the low-temperature phase. In essence, we show how to analytically continue Eq.~\eqref{a3} for negative values of the argument in $\epsilon=4-d$ expansion.

Leaving the details to Appendix \ref{app1}, in $4-\epsilon$ expansion one can rewrite Eq.~\eqref{f} as
\begin{align}\label{PE}
    \sum_{n\neq 0}{\log{\left(1+\frac{\frac{\omega}{\pi}}{n^2}\right)}} & =  -\frac{\omega^{2}}{\epsilon} -\frac{3}{4}\omega ^{2}+\int_{1}^{\infty}dx~\frac{1}{x}(1-e^{-\omega x})\biggl[\vartheta^{4}(x)\nonumber\\
&-1\biggr]+\int_{1}^{\infty}dx~x(1-e^{-\omega x^{-1}})\left[\vartheta^{4}\left( x\right)-1\right]\nonumber\\
&-\biggl[E_{1}\left(\omega\right)+\log{\left(\omega\right)}+\gamma\biggr] -\frac{1}{2}\biggl[1-(1-\omega)e^{-\omega}\nonumber\\
&-\omega^{2}(E_{1}(\omega)+\log{\left(\omega\right)}+\gamma)\biggr]+
O(\epsilon),
\end{align}
where $E_n=\int_1^\infty d\sigma \sigma^{-n}e^{-\sigma z}$ and we have used $E_{3}(\omega)=\frac{1}{2}\left((1-\omega)e^{-\omega}+\omega^{2}E_{1}(\omega)\right)$ in Eq.~$\eqref{NIdr}$.
Using $\lim_{\delta\rightarrow{0^{\pm}}}E_{1}(-|x|\pm i\delta)=-Ei(|x|)\mp i \pi$ and analytically continuing $\log(-|x|)$ such that  $\log(-|x|)=\log(|x|)\pm i\pi$ in upper or lower  half plane respectively, one can write for $\omega<0$
\begin{equation}
    E_{1}(\omega)+\log{\left(\omega\right)} = -Ei(-\omega) +\log{\left(-\omega\right)}.
\end{equation}
Thus for $\omega<0$, Eq.~\eqref{PE} becomes
\begin{align}\label{PE1}
     \sum_{n\neq 0}{\log{\left(1+\frac{\frac{\omega}{\pi}}{n^2}\right)}} & = -\frac{\omega^{2}}{\epsilon} -\frac{3}{4}\omega ^{2}+\int_{1}^{\infty}dx~\frac{1}{x}(1-e^{-\omega x})\biggl[\vartheta^{4}( x)\nonumber\\
&-1\biggr]+\int_{1}^{\infty}dx~x(1-e^{-\omega x^{-1}})\left[\vartheta^{4}\left(x\right)-1\right]\nonumber\\
&-\left[-Ei\left(-\omega\right)+\log{\left(-\omega\right)}+\gamma\right]-\frac{1}{2}\biggl[1-(1-\omega)e^{-\omega}\nonumber\\
& -\omega^{2}(-Ei(-\omega)+\log{\left(-\omega\right)}+\gamma)\biggr]+
O(\epsilon).
\end{align}
The expression \eqref{PE1} can be dimensionally regularized as is done in the case of $\omega>0$ by adding counterterms and removing the pole singularity in $\epsilon$. Notice  that $\int_{1}^{\infty}dx~\frac{1}{x}(1-e^{-\omega x})\left[\vartheta^{4}( x)-1\right]$ diverges for $\omega\leq-\pi$ (even after the dimensional regularization is performed). This is expected since the sum $ \sum_{n\neq 0}{\log{\left(1+\frac{\frac{\omega}{\pi}}{n^2}\right)}}$ itself is not defined for $\omega=-\pi$. Since in practice, from Eq.~\eqref{a3}, the term ${\omega}/{\pi}$ is actually $ \left(2\tilde{x}^2-|\zeta|^{\frac{1}{\nu}}/(4\pi ^2)\right)$, a continuous function of $x$, the sum is only defined for $ \left(2\tilde{x}^2-|\zeta|^{\frac{1}{\nu}}/(4\pi ^2)\right)>-\pi$. The worst case scenario is that of $\tilde{x}=0$, where the argument of the discrete sum is the most negative. Hence for our analytic continuation to hold we must have $-|\zeta|^{\frac{1}{\nu}}<-4 \pi^2$ which translates to $\zeta\gtrapprox-9 $. The expression for $I_{\zeta}(x)$ for $-9<\zeta<0$, can thus be obtained by replacing $\zeta^{1/\nu}$ with $-|\zeta|^{-1/\nu}$ in Eq.~\eqref{a3}.

\begin{figure}[t]
\centering
\includegraphics[width=\linewidth]{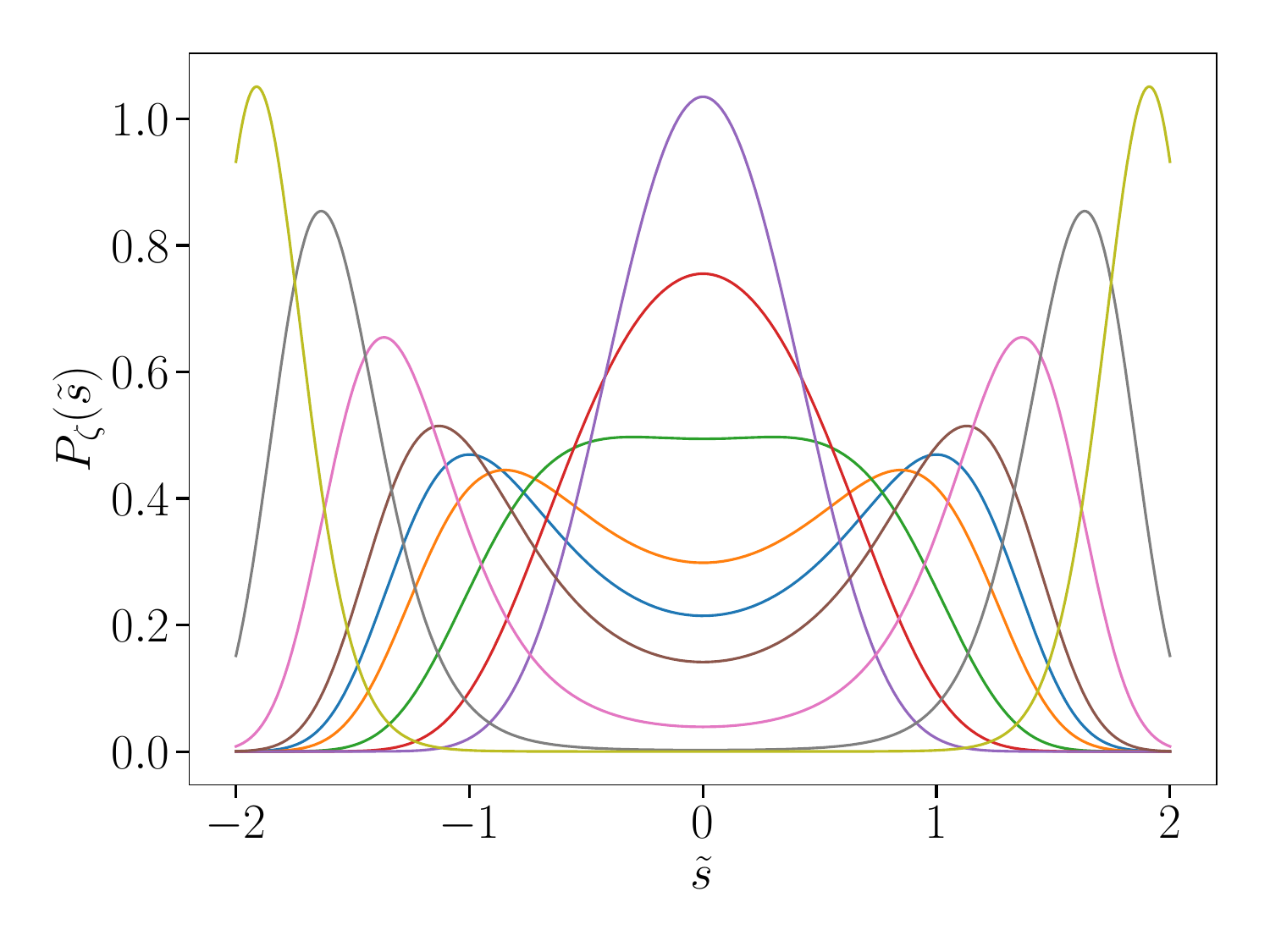}
\caption{ \label{evolution-true-PDF}
Evolution of the PDF obtained at one-loop when $\zeta$ is varied between -4 and 4 in steps of 1 (from bottom to the top curve at the center). The concavity change at the origin occurs for $\zeta_c\simeq 2.12$ at one loop.}
\end{figure}

\begin{figure}[t]
\centering
\includegraphics[width=\linewidth]{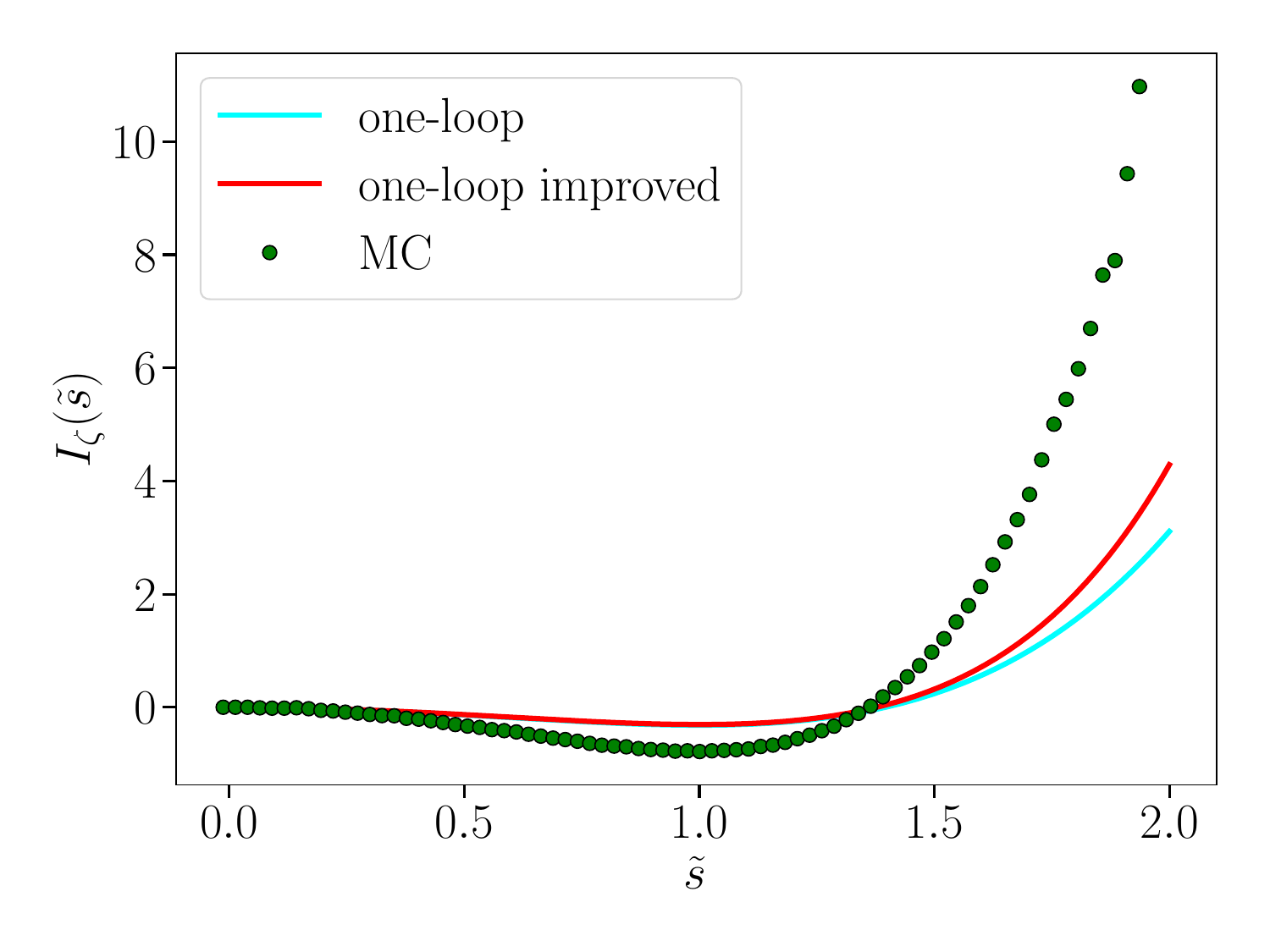}
\caption{ \label{true-zeta=0} Comparison of $I_{\zeta=0}(x)$ obtained either from Monte Carlo (MC) simulations or from improved and unimproved one-loop results. At large field, the improved curve follows almost the same power law as the Monte Carlo data although not with the same prefactor.}
\end{figure}

\begin{figure}[t]
\centering
\includegraphics[width=\linewidth]{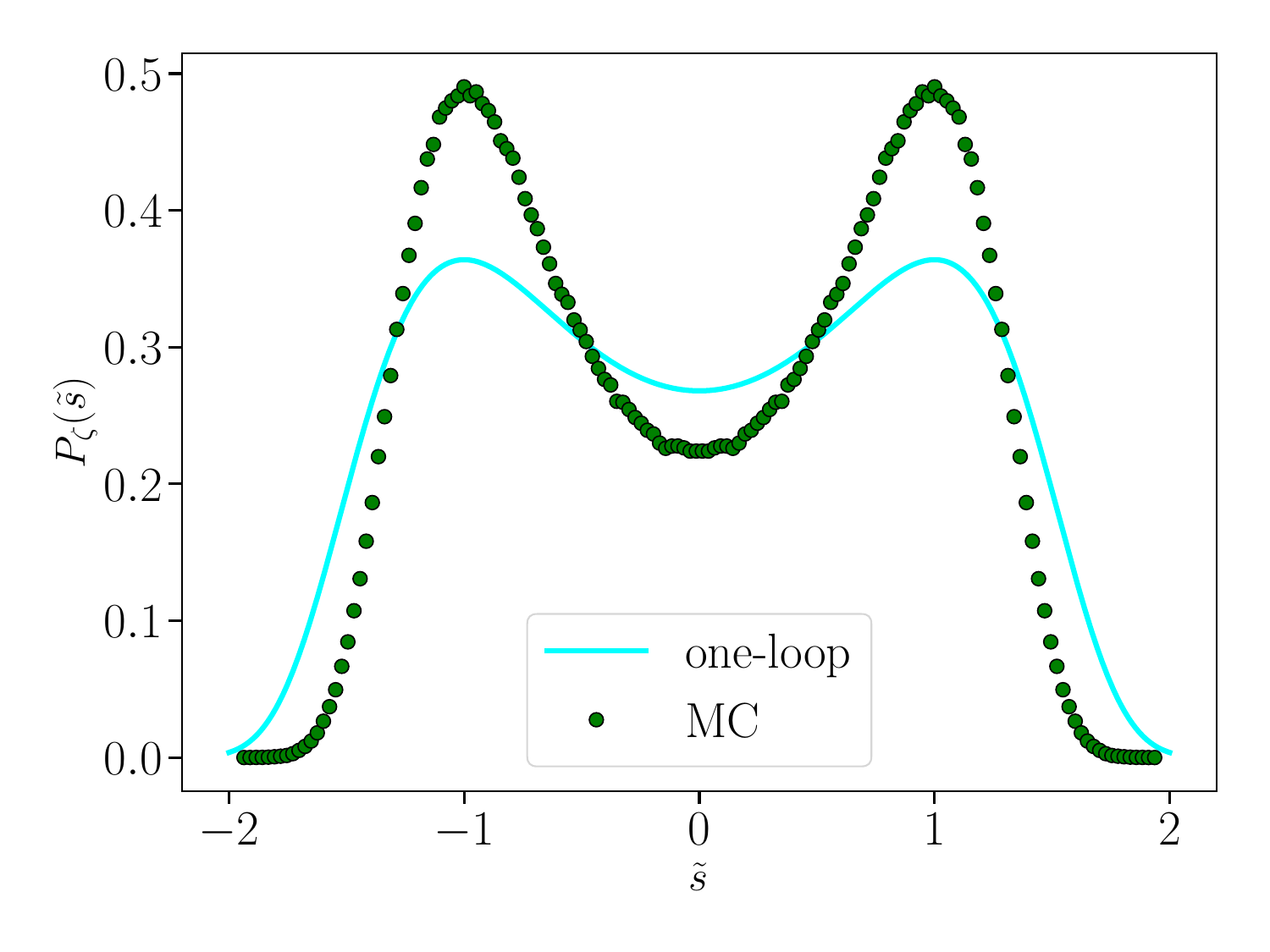}
\caption{ \label{true-PDF-zeta=0} $\zeta=0$: Comparison of the PDF  obtained from Monte Carlo (MC) simulations and from the one-loop expansion. }
\end{figure}

Also, we must point out, that the improvement that works for positive values of $\zeta$ does not work for the negative ones. The reason for this is rather simple. In the low-temperature phase, there exists a non-vanishing magnetization and hence the true large field limit is reached for higher values of $\tilde{x}$. Due to this, even though the large field behavior remains the same for any value of $\zeta$ (positive or negative) i.e. $I_{\zeta}(\tilde{x})\propto\tilde{x}^{{\delta}+1}$ for large $\tilde{x}$, the way to interpolate between the small field and the large field behavior must be different. Thus, getting rid of the large logarithms occurring at large fields is in principle possible, as in the high-temperature phase but the matching with moderate values of the field is less trivial because of the presence of a nonvanishing spontaneous magnetization. 
Designing a family of improvements for negative values of $\zeta$ must therefore be possible but remains beyond the scope of our present work.

\section{Comparison with Monte-Carlo Simulations}\label{MC}
We compare below Monte Carlo (MC) data of Ref.~\cite{Balog2022} and the one-loop results with and without RG improvement, both for the rate function and the PDF. All curves have been obtained in $d=3$ in the Ising case.

The interest in showing these two quantities is that the PDF is of order unity at small fields and extremely small at large fields, making invisible the structure of this function in this field regime, and vice versa for the rate function. In particular, the discussion about the RG improvement will focus only on the rate function.

\subsection{The two-scale-factor universality}\label{two-scale-factor}
The comparison between the PDF obtained from the lattice Ising model and the PDF obtained from the $\phi^4$ model requires a priori to get rid of several non-universal factors.  Once these factors are eliminated, the different PDFs are universal and can be compared. This section aims to analyze how to match results obtained for instance in the lattice Ising model and the $\phi^4$ theory.

The general RG theory tells us that two non-universal amplitudes survive in the renormalized theory: this is called the two-scale-factor universality \cite{Bervillier76}. These two scales are related on one hand to the scale of the order parameter which is of course different between the Ising  model and a scalar field theory and, on the other hand, to the amplitude in front of the parameter $t$ in Eq.~\eqref{hamiltonian} which means that the distance to the critical temperature is {\it a priori} not measured in the same way in the two models. This second amplitude translates to an amplitude for the correlation length when $t$ is eliminated in favor of $\xi_\infty$ and this amplitude is of course also nonuniversal. However, it is possible to decide on a protocol to define in the same way the correlation lengths of the two models. We can, for instance, decide to define the correlation length from the long-distance behavior of the two-point correlation function: $\langle \hat\sigma_i \hat\sigma_j\rangle\propto \exp(-r_{ij}/\xi_\infty)$ for $r_{ij}\to\infty$.\footnote{Note that there are many ways of defining the correlation length, all of which are quantitatively different, but all of which behave like $t^{-\nu}$ close to $T_c$, and are related to each other by a universal amplitude ratio.} Of course, these two correlation lengths are measured in two different units, the lattice spacing $a$ for the Ising model and either the UV cutoff or an arbitrary scale $\mu^{-1}$ in the $\phi^4$ theory such as that introduced in Eq.~\eqref{dimensionless}. Thus, of course, even if these two quantities are defined according to the same definition in the two models, $\xi_\infty/a$ and $\xi_\infty\mu$ are different. However, this discrepancy disappears in the ratio $\zeta$ because here again the system size $L$ is measured in the same unit, either $a$ or $\mu^{-1}$. For instance, if $\zeta=1$ this means in both systems that $\xi_\infty=L$ and the ambiguity in comparing the two systems has disappeared.

The situation is somewhat similar for the scale of the order parameter because $\hat{s}$ does not mean the same thing in the two systems. 
Noting that this scale is independent of $\zeta$, it can be fixed for instance by expressing the PDF as a function of $s/\langle s^2\rangle_{0}$ instead of $s$, where $\langle s^2\rangle_{0}=\int ds s^2 P_L(s,\zeta=0)$ is the variance of the order-parameter at $\zeta=0$.
In this case
\begin{equation}\label{universal-PDF-rescaling}
    \sqrt{\langle s^2\rangle_{0}}\, P_L(s,\zeta)= \tilde{P}\left(\frac{s}{\sqrt{\langle s^2\rangle_{0}}},\zeta\right),
\end{equation}
where $\tilde{P}$ is the truly universal PDF, and the above equality holds only in the large $L$ limit. Since $\sqrt{\langle s^2\rangle_0}\sim L^{(-d+2-\eta)/2}$, the argument of $\tilde{P}$ in Eq.~\eqref{universal-PDF-rescaling} is proportional to $\tilde{s}$ defined in Eq.~\eqref{stilde}.  Thus, to compare two PDFs it is convenient to make a change of variables from $s$ to $s'$ such that $\langle {s'}^2\rangle=1$ \cite{Eisenriegler1987}. In this case,  $P_L(s')$ is directly the universal PDF for $L\to\infty$. The normalization conditions satisfied by the universal PDF are therefore: (i) it is normalized to unity as it must be a probability distribution and, (ii) its second moment is set to unity by a rescaling of the field. 
Alternatively, one can rescale $s$ such that the maxima of the PDF at $\zeta=0$ are at $s=\pm1$. Since the position of the maxima of $\tilde P$ are universal, this amounts to a universal rescaling of the argument of $\tilde P$. It is the latter procedure that we use below.

\subsection{Comparison with the unimproved results}
We show in Fig.~\ref{evolution-true-PDF} how the PDF evolves as a function of $\zeta$.
The first observation we can make when looking at Fig.~\ref{evolution-true-PDF} is that the concavity of the PDF at the origin changes for $\zeta$ between 2 and 3. This means that the PDF changes from a double-peak shape to a single-peak shape.  We have determined the value of the critical value $\zeta_c$ where this change occurs and we have found  $\zeta_c^{\rm 1L}=2.12$ while this occurs for the Monte Carlo simulations for $\zeta_c^{\rm MC}$ between 2.05 and 2.1.\footnote{This bound on $\zeta_c^{\rm MC}$ has been obtained by Monte Carlo simulations as described in \cite{Balog2022}, collecting $\sim 150\times 10^6$ configurations of magnetization and energy for $L=16,32,64$. Using histogram reweighting to change the temperature and thus $\zeta$, we can assert that $\zeta_c^{\rm MC}\in[2.05,2.1]$ independent of the system size. } This shows the very good accuracy of the one-loop determination of this highly nontrivial universal quantity. Let us notice that since the RG improvement is only effective at large fields, the value of $\zeta_c^{\rm 1L}$ is not modified by it.

Second, we observe that for $\zeta>\zeta_c$, the PDF is unimodal and could thus seems Gaussian. This is not too surprising for the typical values of the field because a large $\zeta$ means that $L\gg \xi_\infty\to\infty$. In this case, for the typical values of $\tilde{s}$, the PDF is indeed well approximated by a Gaussian around $\tilde{s}=0$, even at criticality. It is only in the tail of the distribution, that is, in the large field limit that the difference to strict Gaussianity shows up, something almost invisible on the PDF. Furthermore, both numerically and theoretically, the non-Gaussian behavior of the tails can be shown to hold for the rate functions, whose exact leading behavior is $\tilde{s}^{\delta+1}$ \cite{Balog2022}.

\begin{figure}[t]
\centering
\includegraphics[width=\linewidth]{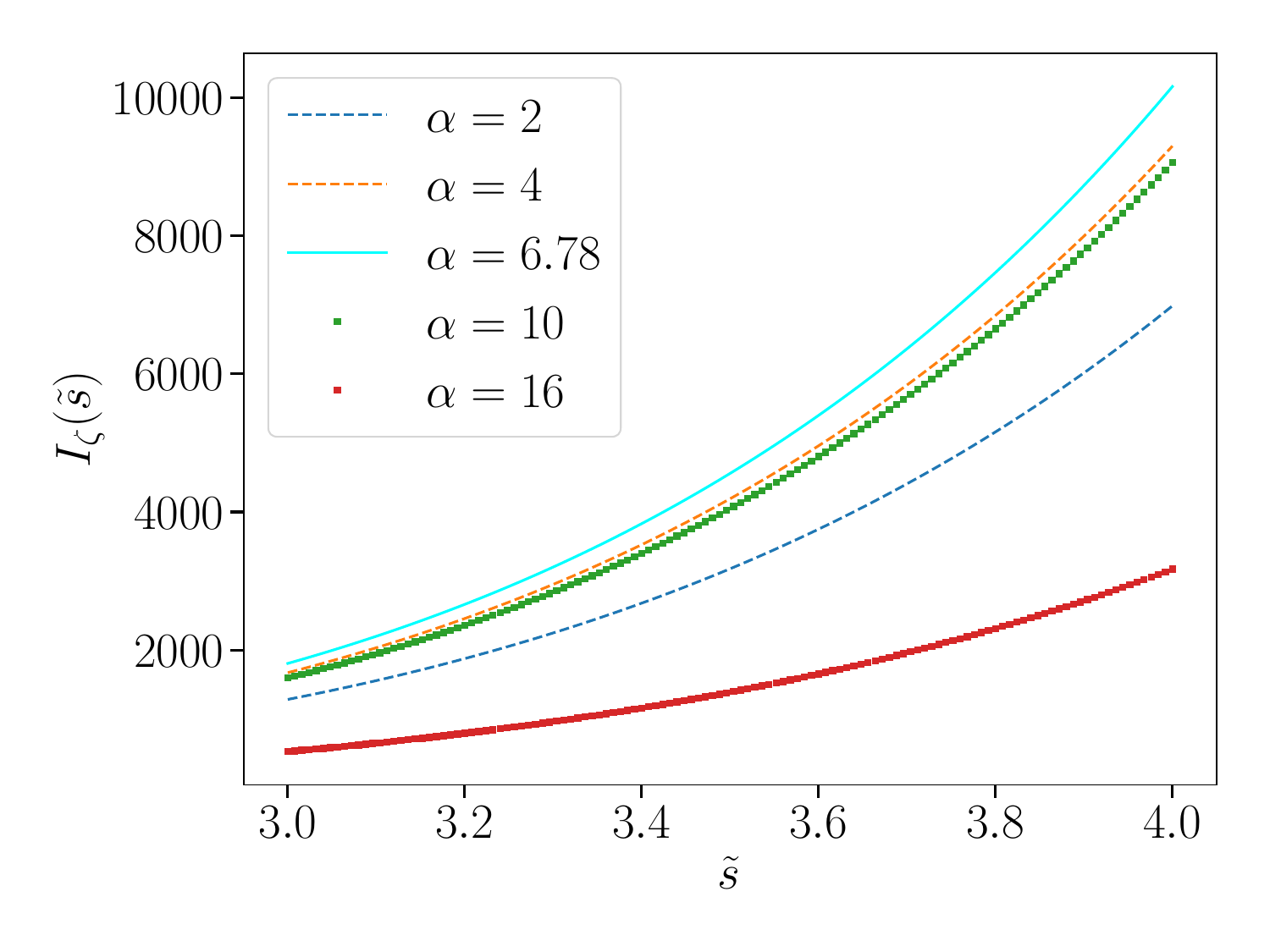}
\caption{\label{pms} Improved rate function at $\zeta=0$ for different values of the parameter $\alpha$ of Eq.~\eqref{IMPE}. By increasing $\alpha$ from 0 to 6.78, the optimum value, all the curves move upwards overall, then, by further increasing $\alpha$, they all move downwards.
}
\end{figure}

We show in Figs.~\ref{true-zeta=0} and \ref{true-PDF-zeta=0} the plots of the rate function and the PDF for $\zeta=0$, where they are compared to the Monte Carlo data of Ref.~\cite{Balog2022}. 
We observe that qualitatively, i.e. for the overall shape, with a single or double peak, the comparison between the PDF or the rate function for $\zeta=0$ with the Monte Carlo data is good but that the quantitative error is rather large. We have checked that typically the same holds true for all positive values of $\zeta$. It is thus rather surprising that the one-loop determination of $\zeta_c$ is so accurate while the overall shape of the PDFs is not.

Let us notice that the comparison of the PDFs at $\zeta=0$ shown in Fig.~\ref{true-PDF-zeta=0} differs from that given in Fig. 1 of Ref.~\cite{Eisenriegler1987}. We agree with the one-loop calculation of the PDF at $\zeta=0$ derived in this reference and that is rederived above. However, we disagree with the fact that the curve obtained by Monte Carlo simulations in Fig.~15 of Ref.~\cite{Binder1981a}, and which is used in Fig. 1 of Ref.~\cite{Eisenriegler1987}, corresponds to $\zeta=0$, as assumed in Ref.~\cite{Eisenriegler1987}. We have found from the data given in Ref.~\cite{Binder1981a} that the value of $\zeta$ to which this PDF corresponds is $\zeta=1.18$. By performing a Monte Carlo simulation for $\zeta=1.18$, we have confirmed the Monte Carlo data given in Ref.~\cite{Binder1981a} and in  Fig.~1 of Ref. \cite{Eisenriegler1987}. Notice also that our Monte Carlo data at $\zeta=0$ agree with the large-scale simulations of Ref.~\cite{Xu2020}. It is therefore a numerical coincidence that the PDF found at one loop for $\zeta=0$ coincides with the PDF found in Monte Carlo simulations for $\zeta=1.18$ and we reaffirm that our Fig.~\ref{true-zeta=0} gives the correct comparison between the one-loop and Monte Carlo results.

As already said in the previous section, the large field behavior is not reproduced by the one-loop results. This is the reason for the RG improvement proposed in Section \ref{improvement} and we show now how it indeed improves the comparison with Monte Carlo data.

\subsection{Comparison with the RG improved results}\label{improved-results}
The reason for the necessity of the improvement and the way the one-loop results can be improved at large field values have been explained in Section \ref{improvement}. What remains to be shown is how the PMS allows us to select an optimal value of $\alpha$ in Eq.~\eqref{IMPE}.

We find that as $\alpha$ increases from $\alpha=0$, the tail of the curve $I_\zeta(x)$ moves globally upwards, then stops moving and by further increasing $\alpha$, finally moves globally downwards. This is shown in Fig.~\ref{pms}. This is the typical scenario where the PMS is useful for optimizing a function because there exists a particular value $\alpha_{\rm opt}$ where the entire function $I_\zeta(x)$ is stationary. We find  $\alpha_{\rm opt}=6.78$.
One crucial point to notice here is that $\alpha_{\rm opt}$ is independent of $\zeta$: it is always for the same value of $\alpha$ that $I_\zeta(x)$ becomes stationary. More about this functional PMS is discussed in Appendix \ref{app2}.

We can see on Fig.~\ref{true-zeta=0} that the tail of the rate function is better reproduced with the improved curve than with the unimproved one. It is in fact possible to show that at large field, the improved curve follows the right power law, that is, $\tilde{s}^{\delta+1}$ with $\delta+1=6$ (instead of 5.79), which is not the case of the unimproved curve. However, the prefactor of this power law is clearly incorrect. The moderate-field region is not well reproduced either. We have checked that this is the case for all values of $\zeta$. We conclude that although the RG improvement does improve the large field behavior, the quantitative discrepancy with the Monte Carlo data persists. This is clearly a consequence of the one-loop approximation.

\begin{figure}[t]
\centering
\includegraphics[width=\linewidth]{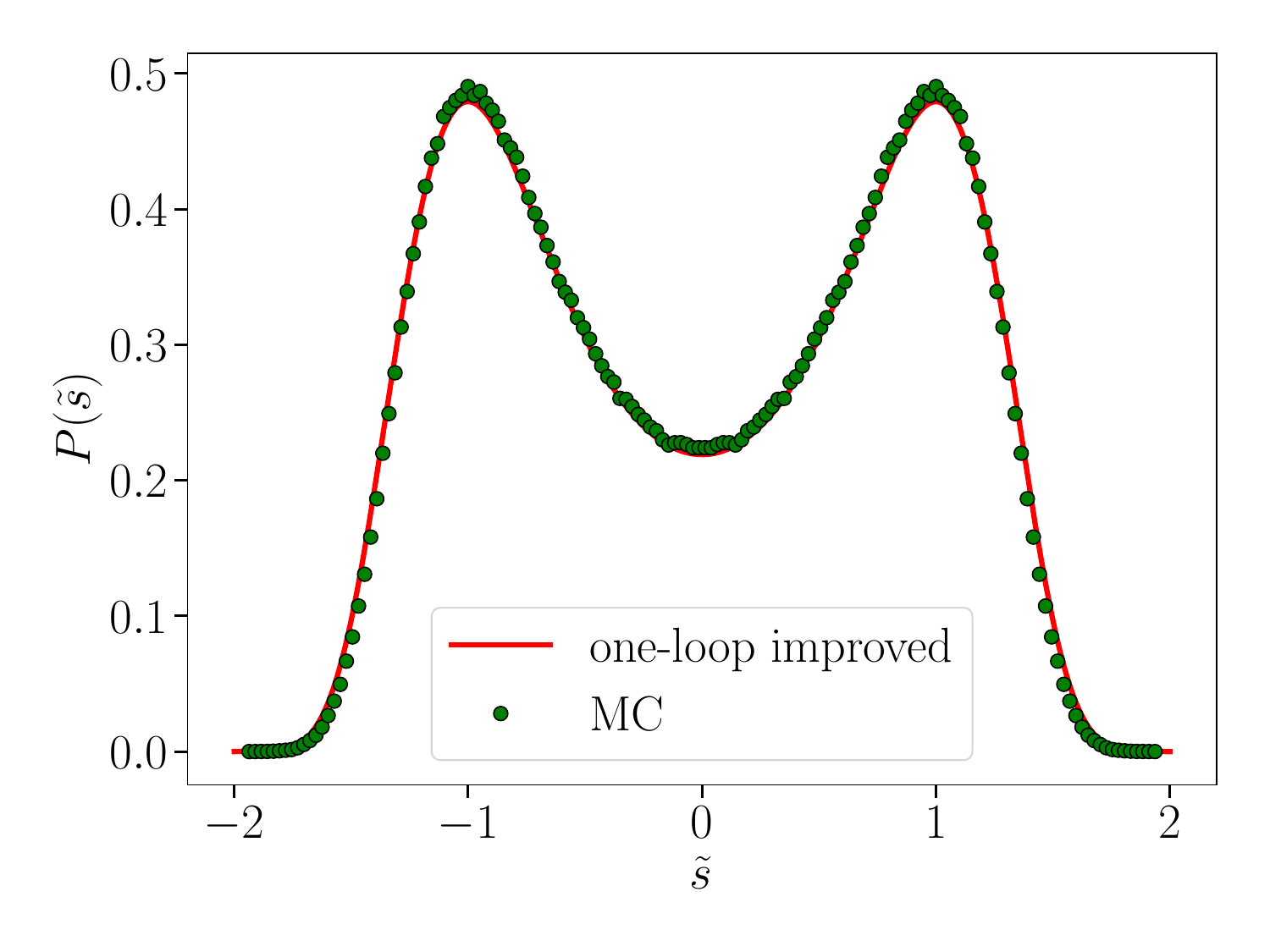}
\caption{\label{pdfzeta=0}
Rescaled PDF $P_{\zeta}^R(\tilde{s})$  for $\zeta=0$ obtained by exponentiating the rescaled  rate function $I_{\zeta=0}^R(\tilde{s})$ , see Fig.~\ref{rescaled-rate-function-zeta=0} and Section \ref{conjecture}.  }
\end{figure}

\begin{figure}[t]
\centering
\includegraphics[width=\linewidth]{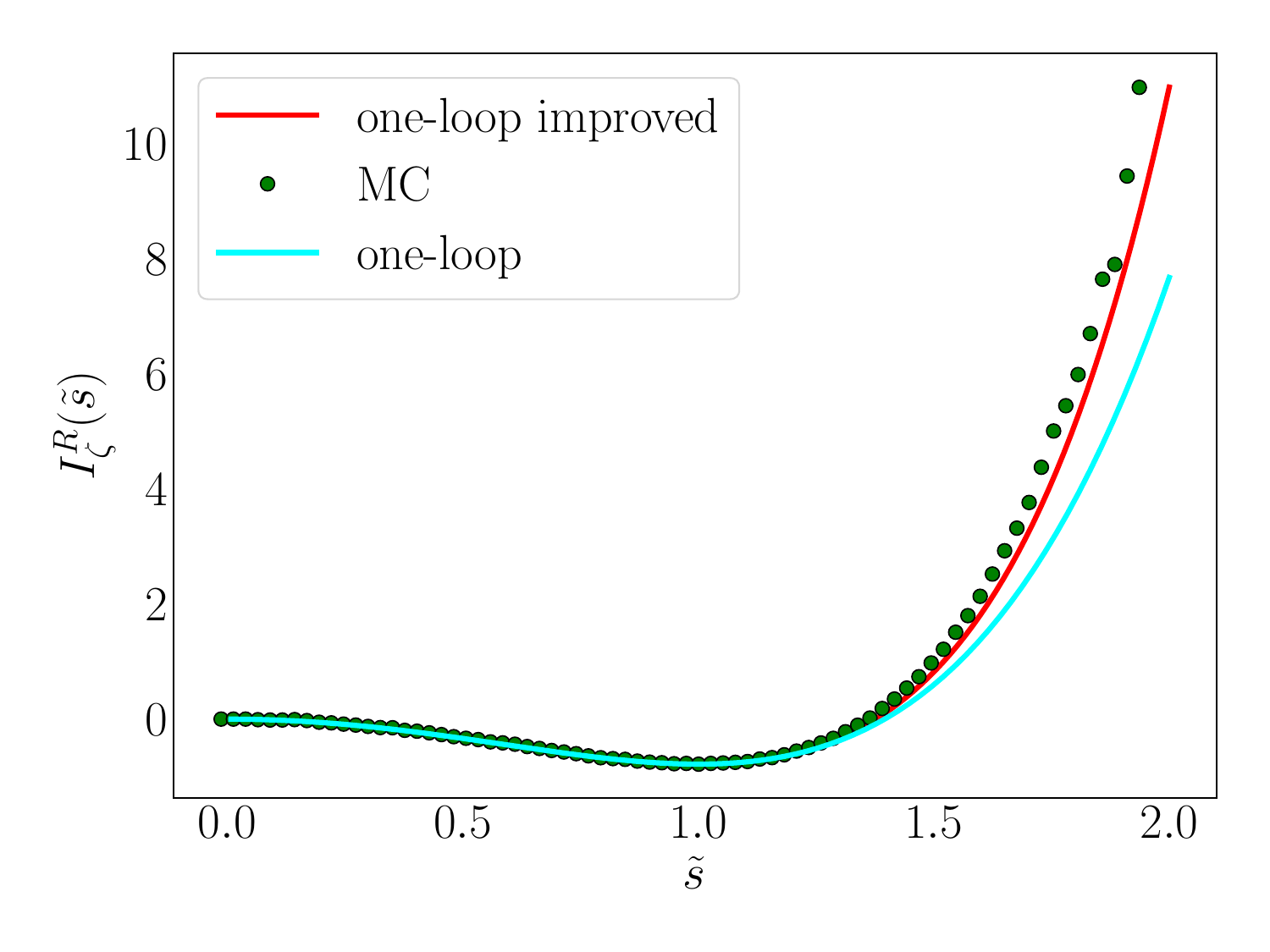}
\caption{\label{rescaled-rate-function-zeta=0}
Rescaled rate function  $I_{\zeta}^R(\tilde{s})$ for $\zeta=0$  (only the one-loop result has been rescaled, not the Monte Carlo data, see Section \ref{conjecture} for a discussion of this rescaling). The one-loop improved curve corresponds to the rate function improved by the choice made in Eq.~\eqref{IMPE} and with the optimal (in the PMS sense) value of $\alpha$: $\alpha_{\rm opt}=6.78$. At large field, the improved curve follows the same power law as the Monte Carlo data.}
\end{figure}

\section{A conjecture about the one-loop results}\label{conjecture}
We would now like to point out an intriguing fact about the one-loop calculation which is worth mentioning.

\begin{figure}[t]
\centering
\includegraphics[width=\linewidth]{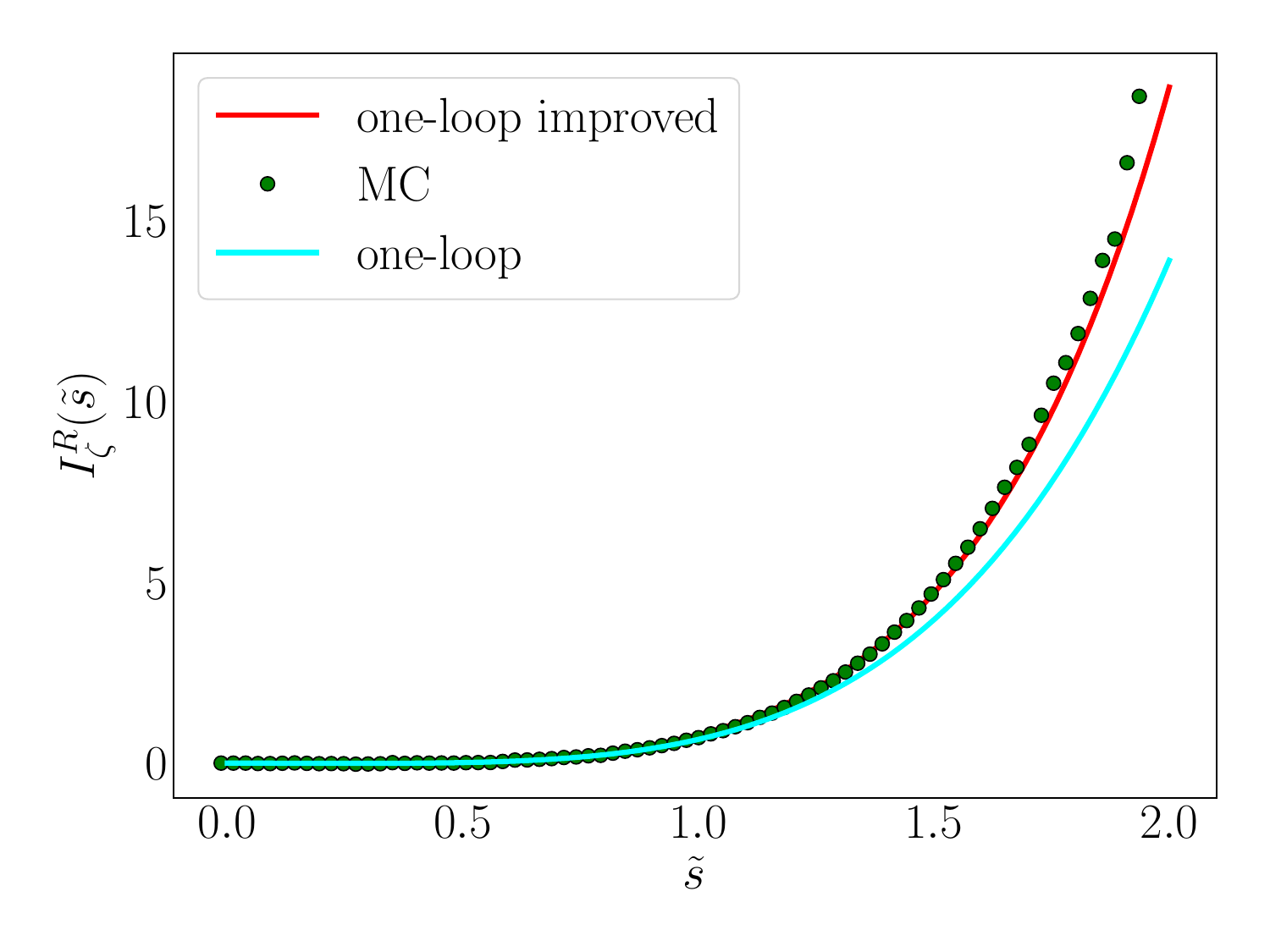}
\caption{\label{rescaled-rate-function-zeta=2}
Same as Fig.~\ref{rescaled-rate-function-zeta=0}  with $\zeta=2$. }
\end{figure}

\begin{figure}[t]
\centering
\includegraphics[width=\linewidth]{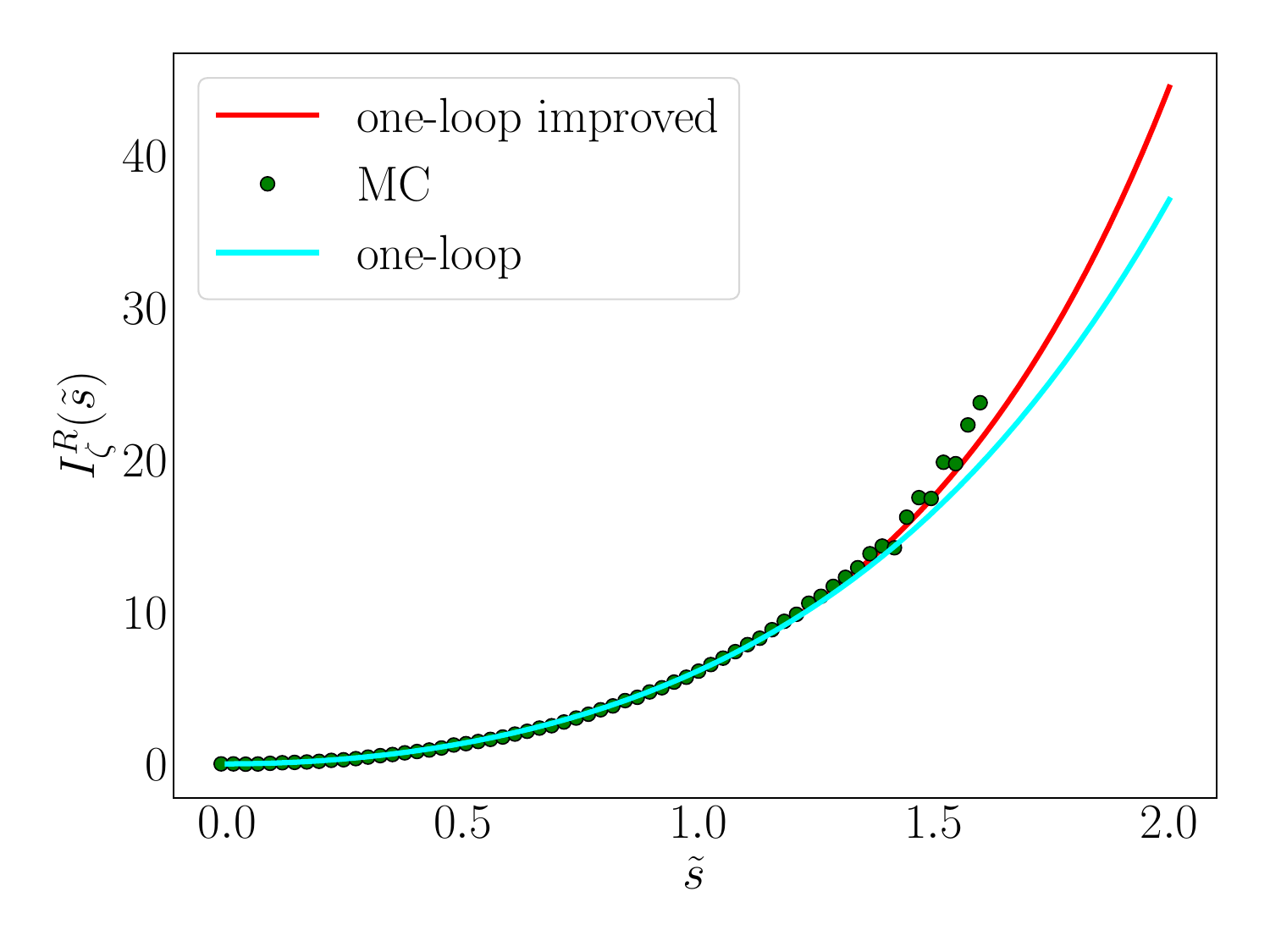}
\caption{\label{rescaled-rate-function-zeta=5}
Same as Fig.~\ref{rescaled-rate-function-zeta=0}  with $\zeta=5$. }
\end{figure}

\begin{figure}[t]
\centering
\includegraphics[width=\linewidth]{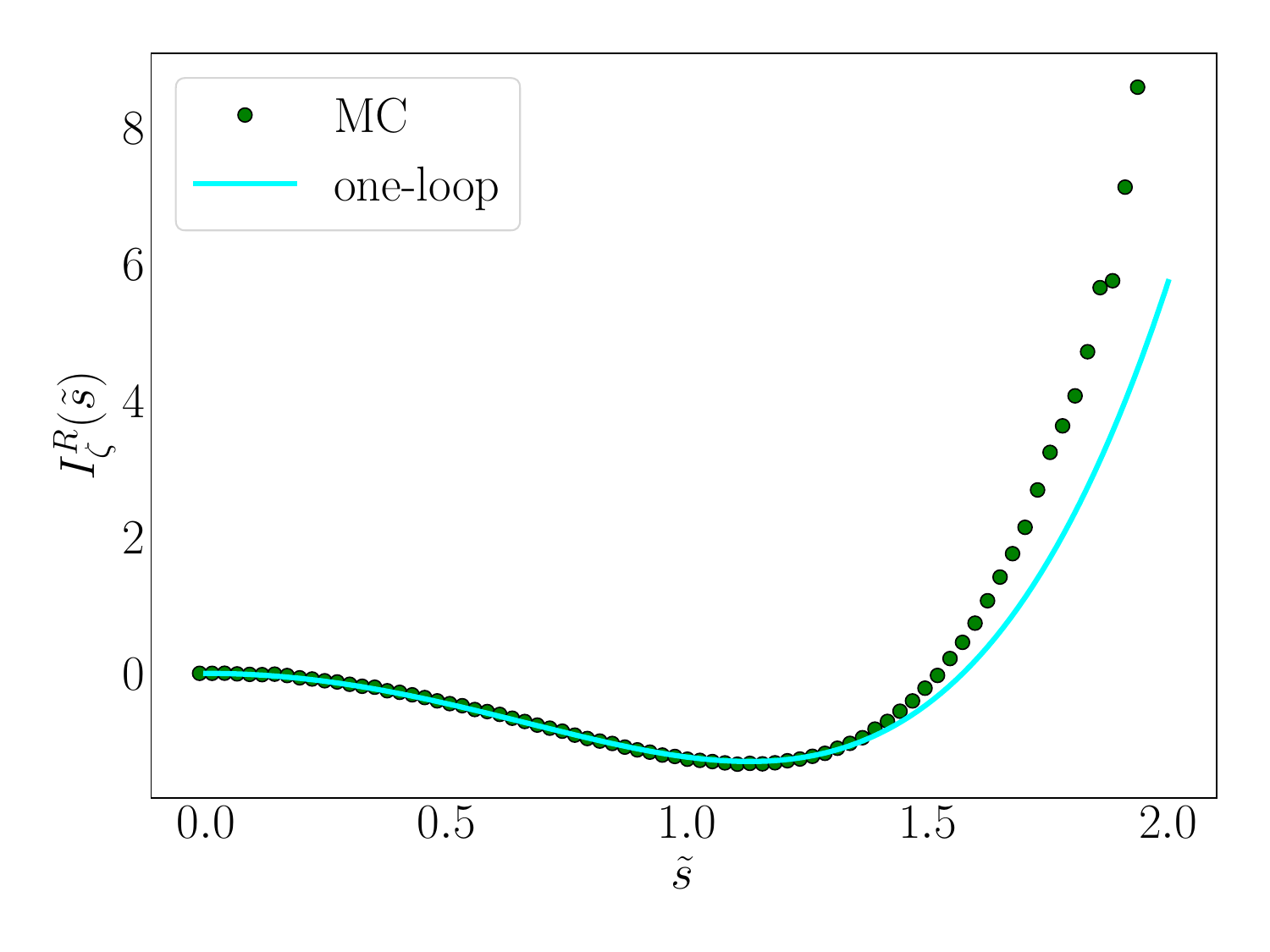}
\caption{\label{rescaled-rate-function-zeta=-1}
Same as Fig.~\ref{rescaled-rate-function-zeta=0}  with $\zeta=-1$. For negative values of $\zeta$, no RG improvement has been implemented.}
\end{figure}

As we have seen in Section \ref{two-scale-factor},  the magnitude of the field should be rescaled to compare the universal PDFs obtained either from Monte Carlo simulations of the Ising model or from the one-loop calculation performed from the $\phi^4$ theory. However, it is not allowed to rescale the rate function itself. If we nevertheless perform such a rescaling for one value of $\zeta$, say $\zeta=0$, and we perform the same rescaling for all positive values of $\zeta$, then we find that there is an almost perfect match between the numerical data and the one-loop results for all values of $\zeta$, see for instance Fig.~\ref{pdfzeta=0} for the case $\zeta=0$. Even more remarkable, we find that the improved one-loop rate functions optimized by the PMS that selects the value of the parameter $\alpha$ defined in Eq. \eqref{IMPE}, that is, $\alpha_{\rm opt.}=6.78$, is the best fit of the numerical data for all values of $\tilde{s}$ and all values of $\zeta$. This turns out to be true also for the analytic continuation to negative values of $\zeta$. In practice, we rescale the one-loop rate function $I_{\zeta}(\tilde{s})$ such that the rescaled rate function $I_{\zeta}^{R}(\tilde{s})$ has the same position and value at its minimum as Monte-Carlo Data at $\zeta=0$. Thus this amounts to two rescalings instead of one which we are permitted to do as discussed in Sec.~\ref{two-scale-factor}.   We show in Figs.~\ref{rescaled-rate-function-zeta=0} to \ref{rescaled-rate-function-zeta=-1} some of these rescaled rate functions $I^{R}_{\zeta}(\tilde{s})$.

Therefore, we conjecture that, unexpectedly, the error from the one-loop approximation in $d=3$, improved at large fields by the choice made in Eq.~\eqref{IMPE} and optimized with the PMS,\footnote{We have tested the robustness of our results by using another family of improvements. Here again, the PMS yields the best possible rate functions compared to the numerical results and for the optimal value of the parameter $\alpha$ of this family these rate functions fit again very well the numerical data for all values of $\zeta$.} is almost entirely concentrated in one number: the scale of the rate functions. If this conjecture is correct,  the one-loop computation has a much stronger predictive power than expected, at least in three dimensions. Obviously, this conjecture needs to be tested on many other models before it can be relied upon.

Note that the rescaling of the rate functions does not alter their concavity at the origin. The value of $\zeta_c$ is therefore unchanged by this rescaling. This was of course a prerequisite of our conjecture, as the value of $\zeta_c$ at one loop is very precise and should hardly vary at two loops and beyond.

\section{Conclusion\label{conclusion}}

In the present paper, we studied the perturbative calculation at one loop of the PDF of the properly normalized sum of spins of an Ising model in dimension $d=4-\epsilon$. We showed that the unique Gaussian fixed point valid for weakly-correlated random variables is replaced at criticality by infinitely many universal PDFs. They depend on how the critical and thermodynamic limits are taken and whether the limit is taken from the high- or low-temperature phase. This infinity of universal PDFs can be parameterized by the ratio $\zeta=L/\xi_\infty$ of the system size to the bulk correlation length in the double limit $L,\xi_\infty\to\infty$ at fixed $\zeta$. While the shape also depends on the boundary conditions, we focused here on periodic boundary conditions. We find that the shape of the PDF changes as $\zeta$ increases and goes from a bimodal shape at small $\zeta$ to a unimodal shape at large $\zeta$. The value $\zeta_c$ where this change occurs is a universal number and it turns out that its one-loop value compares very well with its Monte Carlo value.

While the tails of the PDFs are qualitatively incorrect in the perturbative calculation, it is possible to design RG improvements for these tails that restore the correct leading behavior. Not only is the correct leading behavior restored but also a subleading logarithmic correction is found in agreement with \cite{next_paper_2035}. We show that there are infinitely many such improvements and that using the Principle of Minimal Sensitivity selects one optimal improvement that turns out to be indeed the best improvement when compared to numerical data. However, even after improvement, the comparison between the one-loop and the numerical results remains at a qualitative level. We show that by assuming the existence of one more fit parameter, namely the scale of one rate function for a given value of $\zeta$, it is possible to drastically improve the agreement between the one-loop improved results and the Monte Carlo data. We, therefore, assume that the error from the one-loop approximation is mostly concentrated on one number, the scale of the rate function, and that subsequent orders of perturbation theory simply correct this scale.

One of the questions that naturally arises in connection with our work is that of its range of applicability. For the CLT it is well known: it applies to all systems involving only iid variables having a finite mean and variance. For the $3d$ Ising model, it is clear that all systems belonging to the Ising universality class share the same universal family of PDFs. The question is thus that of the contour of the $3d$ Ising universality class. On one hand, the answer to this question is easy: all $3d$ and $\mathbb{Z}_2$ invariant systems undergoing a second-order phase transition are in the Ising universality class and thus share the same universal PDFs. In this sense, we can speak of a generalization of the CLT. On the other hand, it is difficult to know {\it a priori} whether a given system undergoes a second-order phase transition and the phase diagram of all $\mathbb{Z}_2$ invariant systems in $3d$ is probably impossible to fully characterize. For example, the phase diagram of scalar field theories involving a $\phi^6$ term in their Hamiltonian in addition to the terms already present in the Eq.~\eqref{hamiltonian} is not known because these are theories undergoing either a first-order phase transition or a second-order phase transition with a very non-trivial boundary between the two regions of the parameter space.\footnote{Notice that systems on this boundary can undergo a tricritical phase transition not described by the Wilson-Fisher fixed point but by the Gaussian fixed point.} Therefore, the range of applicability of the above approach is not fully under control but is exactly at the same level as the characterization of a universality class.

Our study paves the way for many others. First, the extension to O$(N)$ models is straightforward at one-loop although the RG improvements must be further studied to take into account the $N$-dependence of the rate functions. It will be interesting to see whether $\zeta_c$ at one loop also agrees with the Monte Carlo data and whether our conjecture works again for $N>1$.  Second, the approach to $d=4$ is interesting for its connection to the triviality problem. Here, the problem is to understand how the large field limit and the  $d\to4$ limit have to be taken since they {\it a priori} do not commute. The approach to criticality from the low temperature phase will  also be interesting since it is probable that the PDFs are again bimodal in $d=4$ for $\zeta<0$.
Third, it should be generalized to universality classes with more complex order parameters such as matrices. Not only will it be interesting to see whether the same phenomenon of PDF shape modification occurs when $\zeta$ is varied, but confirmation of our conjecture will also be an interesting challenge. Of course, the study should also be generalized to non-equilibrium systems undergoing continuous phase transitions or showing generic scaling.

\acknowledgments 
We wish to acknowledge Ivan Balog for discussions and collaboration on this topic.
This work was supported by the Croatian Science fund project HRZZ-IP-10-2022-9423, an IEA CNRS project, and by the “PHC COGITO” program (project number: 49149VE) funded by the French Ministry for Europe and Foreign Affairs, the French Ministry for Higher Education and Research, and The Croatian Ministry of Science and Education.

\appendix
\section{Computing the discrete sum}\label{app1}

We compute the discrete sum that appears in (\ref{e}). We start by considering the sum
\begin{equation}
I_{d}(r)=\frac1{L^d}\sum_{q\neq 0}\frac{1}{r+q^2},
\end{equation}
with $q=2\pi n/L$ and $n\in \mathbb{Z}^d$, which is divergent in the UV. We thus regularize it, and add and subtract a divergent (if not regularized) part $\int \frac{d^d k}{(2\pi)^d}\frac{1}{r+k^2}$, which corresponds to the thermodynamic limit of $I_d(r)$ (i.e. $rL^2\gg 1$).
The now convergent sum reads
\begin{equation}
\begin{split}
    \tilde I_{d}(r)&=\frac1{L^d}\sum_{q\neq 0}\frac{1}{r+q^2}-\int \frac{d^d k}{(2\pi)^d}\frac{1}{r+k^2},\\
            &=\frac{L^2}{4\pi}\int_0^\infty d\sigma e^{-\sigma \frac{L^2r}{4\pi}}\left(L^{-d}\sum_{n\neq 0} e^{-\sigma \pi n^2}-\int \frac{d^dk}{(2\pi)^d} e^{-\sigma \frac{L^2 k^2}{4\pi}}\right).
\end{split}
\end{equation}
Note that now, all integrals and sums are convergent for $r\geq0$. Performing the integral over $k$ and introducing the Jacobi $\vartheta$ function, $\vartheta(\sigma)=\sum_{j=-\infty}^{\infty}e^{-j^2 \pi \sigma}$, we obtain
\begin{equation}
\begin{split}
    \tilde I_{d}(r) &=\frac{L^{2-d}}{4\pi}\int_0^\infty d\sigma e^{-\sigma L^2r/4\pi}\left(\vartheta^d( \sigma)-1-\sigma^{-d/2}\right).
\end{split}
\end{equation}
This expression is well defined both in the IR ($\sigma\to\infty$, for $d>2$) and the UV ($\sigma\to0$, for $d\leq4$) thanks to the property $\vartheta(\sigma)=\sqrt{\frac{1}{\sigma}}\vartheta(1/\sigma)$ shown by Poisson summation.

Integrating $I_d(r)$ with respect to $r$, we obtain
\begin{equation}\label{Id}
\frac{1}{L^d}\sum_{q\neq 0}\ln\left(1+\frac{r}{q^2}\right)=\frac{1}{L^d}\left(\theta(L^2r/4\pi)-\theta(0)\right)+\int \frac{d^dk}{(2\pi)^d} \ln\left(1+\frac{r}{k^2}\right),
\end{equation}
with 
\begin{equation}
\theta(z)=-\int_0^\infty d\sigma \frac{e^{-\sigma z}}\sigma\left(\vartheta^d(\sigma)-1-\sigma^{-d/2}\right).
\end{equation}

To perform the analytic continuation to negative $z$, it is convenient to recast $\theta(z)-\theta(0)$ as follows,
\begin{align}\label{theta}
    \theta(z)-\theta(0)  =&\lim_{\delta\rightarrow 0} \biggl(\int_{\delta}^{\infty}\frac{d\sigma}{\sigma}\left(\vartheta^{d}(\sigma)-1-\sigma^{-d/2}\right)\nonumber\\
 & - \int_{\delta}^{\infty}\frac{e^{-z\sigma}}{\sigma}\left(\vartheta^{d}(\sigma)-1-\sigma^{-d/2}\right)\biggr)\nonumber\\
     = & \lim_{\delta\rightarrow 0}\biggl(\int_{\delta}^{1}\frac{d\sigma}{\sigma}\left(\vartheta^{d}(\sigma)-1\right)\left(1-e^{-z\sigma}\right)\nonumber\\
   &+\int_{1}^{\infty} \frac{d\sigma}{\sigma}\left(\vartheta^{d}(\sigma)-1\right)\left(1-e^{-z\sigma}\right)\nonumber\\
& -\int_{\delta}^{\infty} ds\left(1-e^{-z\sigma}\right)\sigma^{-d/2-1}\biggr),
 \end{align} 
 where we have regularized the lower bound of the integral to ensure convergence during the various manipulations.
Making the change of variables $\sigma\rightarrow \frac{1}{\sigma}$ in the first integral that appears in the expression \eqref{theta} and once again using the identity $\vartheta(\sigma)=\sqrt{\frac{1}{\sigma}}\vartheta(1/\sigma)$, we get
\begin{align}
\theta(z)-\theta(0)
  & =  \lim_{\delta\rightarrow 0}\biggl(\int_{1}^{\frac{1}{\delta}}d\sigma~~\sigma^{d/2-1}\left(\vartheta^{d}\left(\sigma\right)-1\right)\left(1-e^{-z\sigma^{-1}}\right)\nonumber\\
  &+\int_{1}^{\infty} \frac{d\sigma}{\sigma}\left(\vartheta^{d}(\sigma)-1\right)\left(1-e^{-z\sigma }\right)\nonumber\\
    &- \int_{1}^{\infty}d\sigma~(1-e^{-z\sigma})\sigma^{-d/2-1}-\int_{\delta}^{1}\frac{d\sigma}{\sigma}(1-e^{-z\sigma})\biggr).
\end{align}
This allows us to take the limit $\delta\rightarrow 0$ safely. Using the relations $E_{n}(z)=\int_{1}^{\infty}d\sigma\frac{e^{-z\sigma}}{\sigma^{n}}$ and $E_{1}(z)+\log(z)+\gamma=\int_{0}^{1}\frac{d\sigma}{\sigma}(1-e^{-z\sigma})$, we obtain
  \begin{align}
       \theta(z)-\theta(0)   
       & =\int_{1}^{\infty}d\sigma~~\sigma^{d/2-1}\left(\vartheta^{d}\left(\sigma\right)-1\right)\left(1-e^{-z\sigma^{-1}}\right)\nonumber\\
       &+\int_{1}^{\infty} \frac{d\sigma}{\sigma}\left(\vartheta^{d}(\sigma)-1\right)\left(1-e^{-z\sigma}\right)\nonumber\\
       &-(E_{1}(z)+\log{z}+\gamma)-\left[\frac{2}{d}-E_{\frac{d}{2}+1}(z)\right].
 \end{align}
Defining $\omega=\frac{rL^2}{4\pi}$, and performing the change of variable $k\rightarrow \frac{kL}{2\pi}$, Eq. \eqref{Id} becomes 
\begin{align}\label{NIdr}
    \sum_{n\neq 0}\ln\left(1+\frac{\frac{\omega}{\pi}}{n^2}\right) & = \int d^{d} k\ln\left(1+\frac{\frac{\omega}{\pi}}{k^2}\right)\nonumber\\
    & + \int_{1}^{\infty}d\sigma~~\sigma^{d/2-1}\left(\vartheta^{d}\left(\sigma\right)-1\right)\left(1-e^{-\omega \sigma^{-1}}\right)\nonumber\\
     &+\int_{1}^{\infty} \frac{d\sigma}{\sigma}\left(\vartheta^{d}(\sigma)-1\right)\left(1-e^{-\omega \sigma}\right)\nonumber\\
       &-(E_{1}(\omega)+\log{\omega}+\gamma)-\left[\frac{2}{d}-E_{\frac{d}{2}+1}(\omega)\right].
\end{align}

\begin{figure}[t]
\centering
\includegraphics[width=0.8\linewidth]{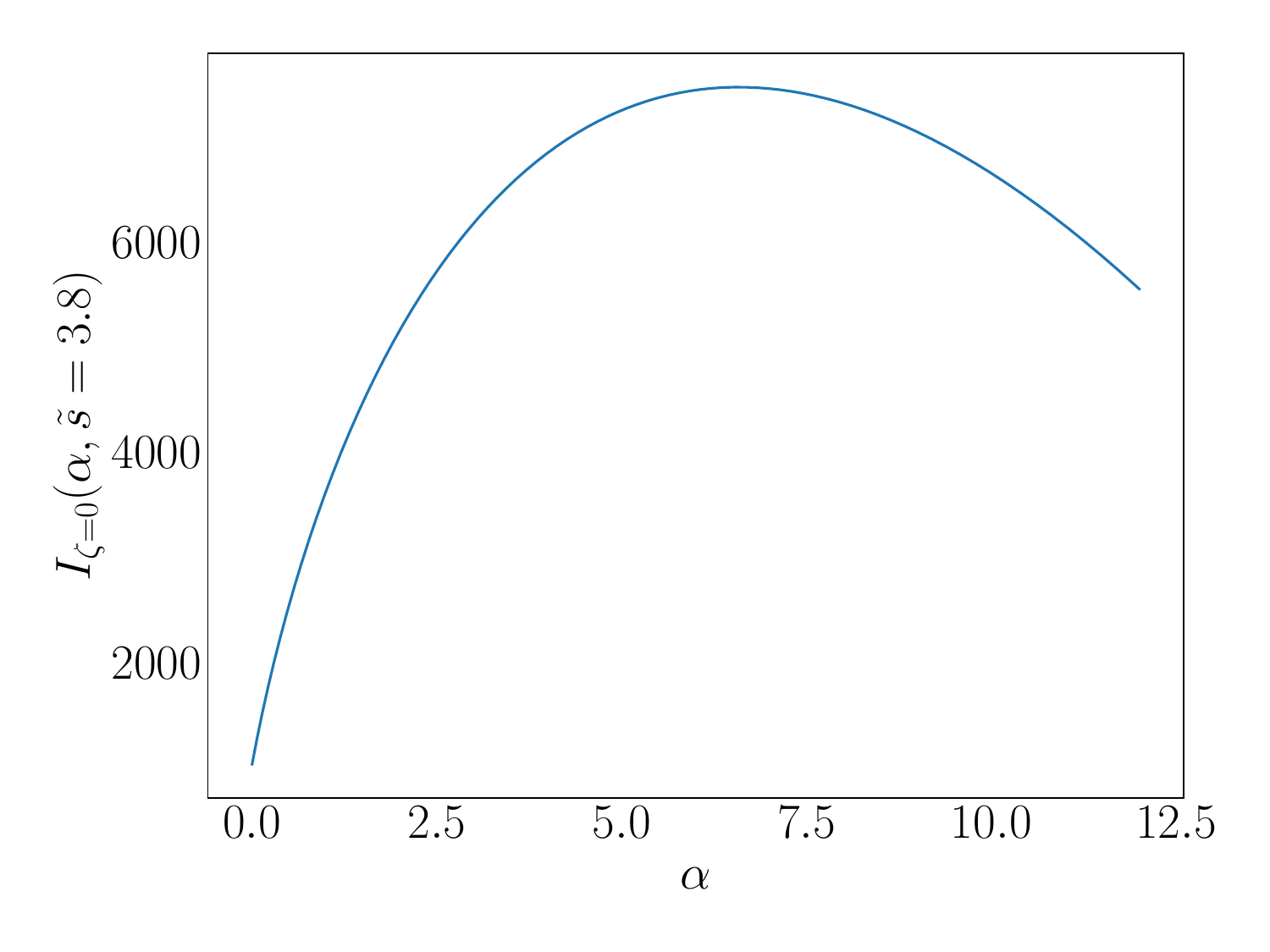}
\caption{\label{PMS-Max} $I_{\zeta=0}(\tilde s)$ plotted at $x=3.8$ for varying values of $\alpha$.}
\end{figure}

\section{The Functional PMS \label{app2}}

In Fig.~\ref{pms}, we show that as the parameter $\alpha$ of the functional optimization, see Eq.~\eqref{IMPE}, is increased,  $I_{\zeta}(\tilde s)$ first moves upward then reaches a stationary curve at $\alpha_{\rm opt}=6.78$  before starting to move downward. A key point to note here is that this phenomenon happens for all values of $\zeta$, and with the same value $\alpha_{\rm opt}$.

\begin{figure}[h]
\centering
\includegraphics[width=0.8\linewidth]{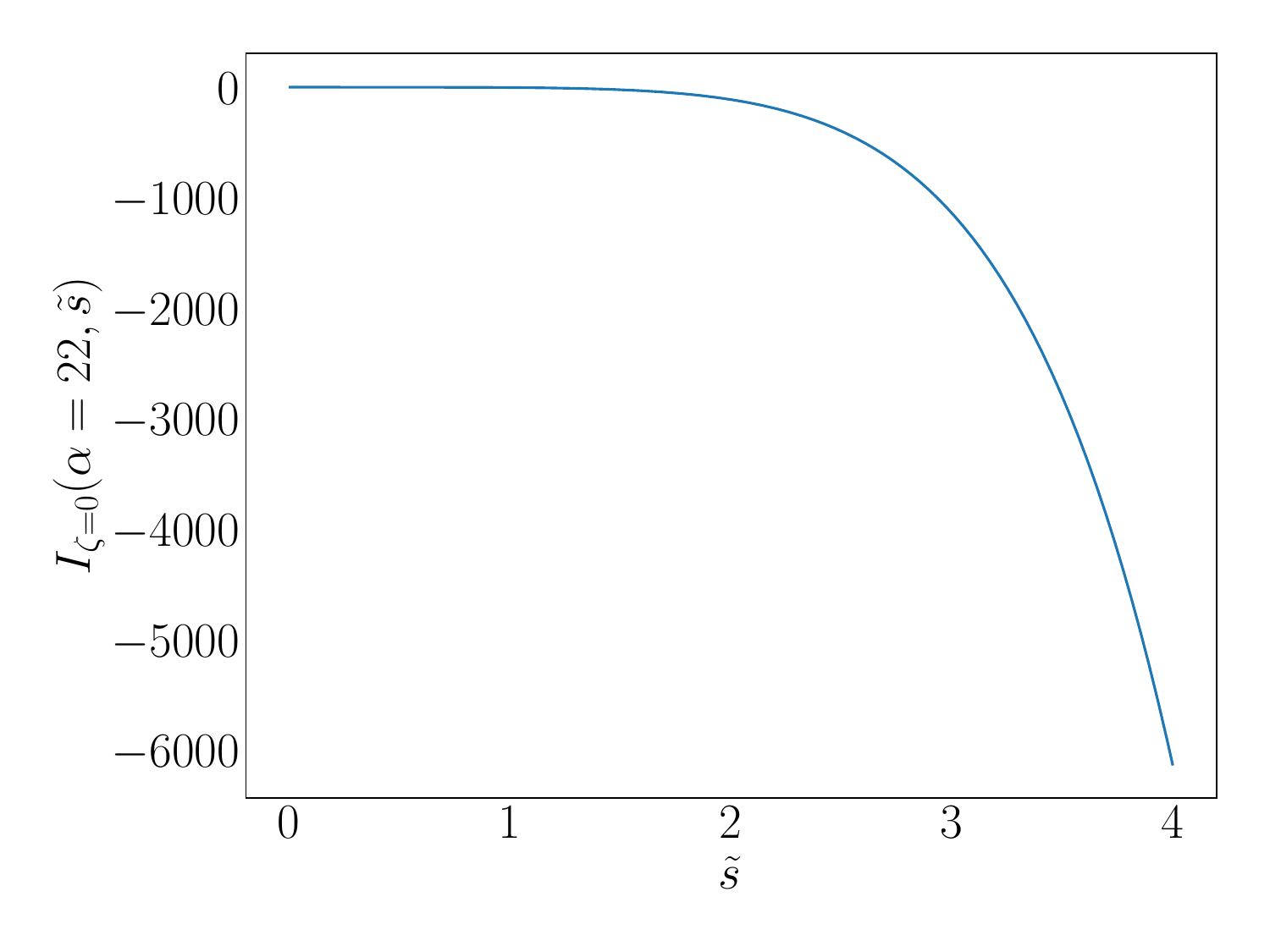}
\caption{\label{PMS-fail} $I_{\zeta=0}(\tilde s)$ for $\alpha=22$.}
\end{figure}

A close examination of what happens, value of $\tilde s$ by value of $\tilde s$, shows that the value of $\alpha$ where $I_{\zeta}(\alpha,\tilde s)$ is stationary depends slightly on $\tilde s$: $\alpha_{\rm opt}=\alpha_{\rm opt}(\tilde s)$. However, this dependence on $\tilde s$ is very weak and we find that for $\tilde s>2.5$,  $\alpha_{\rm opt}(\tilde s)$ varies between 6.31 and 6.78, see the value of $\alpha_{\rm opt}(\tilde s=3.8)=6.58$ in Fig.~\ref{PMS-Max}. We have plotted $I_{\zeta}(\alpha,\tilde{s})$ for $\alpha=6.31$ and $\alpha=6.78$ and found that these curves are almost on top of each other. This also holds true for all values of $\zeta$.
Notice that if we keep on increasing the value of $\alpha$, the rate function becomes a strictly decreasing function as can be seen in Fig.~\ref{PMS-fail}, suggesting an unphysical choice of $\alpha$.

\bibliography{biblio}
\bibliographystyle{apsrev4-1} 

\end{document}